\documentclass{emulateapj}
\usepackage{graphicx}
\begin{document}

\def\etal{et al.\ \rm}

\title{Atmospheres of protoplanetary cores: critical 
mass for nucleated instability.}

\author{R. R. Rafikov}
\affil{IAS, Einstein Dr., Princeton, NJ 08540}
\email{rrr@ias.edu}


\begin{abstract}
Understanding atmospheres of protoplanetary
cores is crucial for determining the conditions
under which giant planets can form by nucleated instability. 
We systematically study quasi-static atmospheres of accreting 
protoplanetary cores for different opacity behaviors and 
realistic planetesimal accretion rates in various parts of 
protoplanetary nebula. We demonstrate that there 
are two important classes of atmospheres: (1) those having outer 
convective zone which smoothly merges with the surrounding 
nebular gas, and (2) those possessing almost isothermal 
outer radiative region which effectively decouples atmospheric 
interior from the nebula. A specific type of gaseous envelope 
which accumulates around a given core depends only 
on the relations between the Bondi radius of the core, photon 
mean free path in the nebular gas, and the luminosity radius 
(roughly the size of the sphere which can radiate accretion 
luminosity of the core at an effective temperature 
equal to the local nebular temperature). Cores in the inner 
parts of protoplanetary disk (within roughly 0.3 AU from the 
Sun) have large 
luminosity radii resulting in the atmospheres of the first 
type, while cores in the giant planet region (beyond several AU) 
have small luminosity radii and always accumulate massive 
atmospheres of the second type. Critical core mass needed 
for the nucleated instability to commence is found to vary 
considerably as a function of distance from the Sun. This mass 
is $5-20$ M$_\oplus$ at $0.1-1$ AU which is too large to permit 
the formation of ``hot Jupiters'' by nucleated instability 
near the cores that have grown in situ. In the region 
of giant planets critical core mass depends on 
the gas opacity and planetesimal accretion rate but is 
insensitive to the nebular temperature or density 
provided that the opacity in the outer radiative region does not
depend on the gas density (e.g. dust opacity). This is true
irrespective of whether the envelope's interior is
convective or radiative and numerical values of the critical
core mass are similar in the two cases. Critical mass in the 
region of giant planets can be as high as $20-60$
M$_\oplus$ (for opacity $0.1$ cm$^2$ g$^{-1}$) if planetesimal 
accretion is fast enough for protoplanetary cores to form prior 
to the nebular gas dissipation. This might 
indicate that giant planets in the Solar System have 
gained massive gaseous atmospheres by nucleated instability  
only after their cores have accumulated most of the mass in solids 
during the epoch of oligarchic growth, 
subsequent to which planetesimal accretion slowed down and 
cores became supercritical. 
\end{abstract}



\section{Introduction.}\label{sect:intro}


Recent discoveries of Jupiter-like planets around
other stars have boosted up efforts to understand the 
origin of giant planets, subject which has been under scrutiny since 
sixties (Safronov 1969; see Brush 1990
for a historical perspective). Currently one of the most 
popular and successful theories of giant planet formation
is a core (or nucleated) instability hypothesis 
(Harris 1978; Mizuno \etal 1978). According to 
this idea massive hydrogen/helium atmospheres of planets like Jupiter 
and Saturn have been acquired as a result of unstable gas
accretion onto a preexisting core made of rock and/or 
ice. Analytical arguments (Stevenson 1982; Wuchterl 1993) and 
numerical calculations (Mizuno 1980; Ikoma \etal 2000)
suggest that the onset of core instability occurs when the mass
of the gaseous atmosphere around the core
becomes comparable to the core mass itself.

The following gedankenexperiment can help understand the 
nature of this instability: imagine placing a massive solid 
core into an infinite homogeneous gaseous 
medium. Gravitational pull from the core gives rise to a 
significant pressure perturbation near the core if the 
escape speed from its surface is larger than the gas 
sound speed, see \S \ref{sect:length_scales}. On a 
short (dynamical) timescale gas settles into
a pressure equilibrium configuration near the core. Since the 
thermal timescale is usually longer than the dynamical timescale, 
this gaseous envelope\footnote{We use terms ``envelope'' and 
``atmosphere'' interchangingly.} initially has entropy 
equal to that of the gas in the surrounding nebula.
Gas temperature at the core surface is rather high 
and this drives an outward transport of energy causing entropy 
of the envelope to decrease. As a result, gas density 
near the core goes up and  envelope becomes more and more massive 
by slow accumulation (on a thermal timescale) of gas 
from the surrounding nebula. 

Given enough time, atmosphere around the core that is 
not too massive cools down to the nebular temperature and 
becomes isothermal, acquiring large gaseous mass.
Such atmosphere has exponential density 
profile with most of the mass concentrated near the core 
surface (Sasaki 1989). Cores are "not too massive" when the mass of this
isothermal atmosphere is much smaller than the core mass.
At the same time, exponential sensitivity to the core gravity 
(and mass) causes atmospheric mass to increase faster than 
the first power of the core mass. Because of that, different fate 
awaits gas around the core that is 
more massive: as the atmosphere cools down  
its mass at some point becomes comparable to the core mass 
and gas starts contributing significantly to the planetary gravity. 
Hydrostatic equilibrium cannot be established beyond this 
point, because accretion of more gas, which helps to 
reestablish pressure equilibrium, also acts now to  
increase the gravitational acceleration. As a result, 
instability commences allowing gas to accrete 
rapidly onto the core. Critical core mass at which 
instability becomes possible is thus set by the condition 
that the mass of the equilibrium atmosphere around the critical 
core is comparable to the core mass itself [apart from
additional logarithmic factors intrinsic of the isothermal case, 
see Sasaki (1989)]. Note that if 
the core is very massive, or gas around it is very dense, then
even the mass of the isentropic envelope forming around the core
on the dynamical time after the core has been introduced into the 
nebula might exceed the core mass, making atmosphere unstable
from the very start.

Real protoplanetary cores are not likely
to be just passive solid bodies of constant mass. 
They accrete planetesimals and accretion not 
only increases the core mass with time but also 
leads to the release of substantial amount of energy at
the surface of the core. This changes the steady state 
structure of the envelope which can no longer be purely 
isothermal (as it tends to become in the case of passive 
core).  Even if we forget for a moment 
about the increase of the core mass caused 
by planetesimal accretion\footnote{Energy release at the 
core surface might also come from radioactive decay or 
differentiation in the core which are not accompanied by 
the change in the core mass.}, luminosity coming from the 
core would cause temperature in a steady state atmosphere 
to be higher near the core surface than in the nebula. As 
a result,  the gas at the same pressure 
is less dense around the radiating core than around 
passive one, meaning that the total atmospheric mass is 
{\it lower} for luminous core than for passive one. 
Nevertheless, similar to the isothermal case, the more 
massive the core is, the more massive is the envelope around it, 
and scaling between them is faster than linear. As a result,
a concept of critical core mass holds again:
stable equilibrium  atmospheres cannot exist around the 
cores which are so massive that their atmospheric mass 
exceeds the core mass.

Previous numerical and analytical studies confirm this 
general picture of the core instability, but there are still 
some open issues. For example, Perri \& Cameron (1974)
and later Wuchterl (1993) have concluded that protoplanetary 
atmospheres are likely to be convective and that their interior 
structure and mass (which was found to be quite low) 
sensitively depend on the density and temperature of the gas
in the surrounding nebula. Stevenson (1982) arrived at a very 
different conclusion by considering a simple analytical 
model of an atmosphere with a constant opacity. He found it 
to be radiative, massive, and its structure to be  
insensitive to the external conditions. His findings confirmed 
previous numerical results by Harris (1978), Mizuno \etal 
(1978), and Mizuno (1980).

The purpose of this study is to systematically explore the 
structure of atmospheres around accreting protoplanetary 
cores under a variety of conditions typical in the protoplanetary
nebulae. Using analytically tractable but still realistic
models we attempt to resolve the aforementioned issues 
concerning the state of the envelope. We
try to single out physical parameters crucial for determining 
the atmospheric structure and this provides us with a general 
classification scheme of possible protoplanetary atmospheres.
We are looking for the steady state structures under the 
explicit assumption that the mass of the gaseous envelope 
is smaller than the mass of the core which it surrounds. 
This restricts our quantitative results to cores with 
masses below critical mass, although all qualitative 
conclusions are valid even for critical cores and we 
use this to estimate the masses of critical cores, 
see \S \ref{subsect:crit_mass}. 

We start by laying down the basics of the problem at hand  
 --- equations, important length and mass scales, 
boundary conditions --- in \S \ref{sect:setup}. In \S 
\ref{sect:solutions} \& \ref{sect:solutions_inner} we derive  
solutions for two important classes of envelope structures 
typical in the region of giant 
and terrestrial planets. Envelope masses and critical core
mass are calculated in \S \ref{subsect:mass}. Our results 
and their implications are discussed in \S \ref{sect:disc}. Finally, 
we devote Appendices to technical issues which emerge in
our calculations.


\section{Problem setup.}\label{sect:setup}


Throughout this study the following approximation 
to the protoplanetary disk structure [similar to 
the Minimum Mass Solar Nebula (MMSN)] is used:
\begin{eqnarray}
&& \Sigma_g(a)\approx100\Sigma_p(a)\approx 
3000~\mbox{g cm}^{-2}~a_1^{-3/2},\\
&& T_0(a)\approx 300~\mbox{K}~a_1^{-1/2},
\label{eq:MMSN} 
\end{eqnarray}
where $\Sigma_p, \Sigma_g$ are the particulate and gas surface 
densities correspondingly, $T_0$ is the gas temperature, 
and $a_n\equiv a/(n~\mbox{AU})$
is a distance from the Sun $a$ scaled by $n$ AU.  
From (\ref{eq:MMSN}) one can find the gas sound speed $c_0\equiv 
(kT_0/\mu)^{1/2}$ ($k$ is a Boltzmann constant and $\mu$ is a 
mean molecular weight) and gas density at 
the midplane $\rho_0\equiv \Sigma_g\Omega/c_0$ 
[$\Omega\equiv (GM_\odot/a^3)^{1/2}$ is the orbital angular frequency,
$M_\odot$ is the Solar mass]:
\begin{eqnarray}
&& c_0(a)\approx 10^5~\mbox{cm s}^{-1}~a_1^{-1/4},\\
&& \rho_0(a)\approx 6\times 10^{-9}~\mbox{g cm}^{-3}~a_1^{-11/4}.
\label{eq:MMSN_deriv} 
\end{eqnarray}
All numerical estimates in the text refer to this 
particular model of the protoplanetary disk.


\subsection{Basic equations.}\label{sect:struct}

Our purpose is to calculate a spherically symmetric distribution of gas density 
$\rho$, pressure $P$, and temperature $T$ around a solid core with 
the mass $M_c$ embedded in the nebular gas. 
Spherical symmetry requires nebula to be 
at least roughly homogeneous around the core and we determine 
the conditions for this to be true in the next section.
Structure of the static envelope as a function of 
distance $r$ from the center of the core is governed by the
equation of hydrostatic equilibrium:
\begin{eqnarray}
\frac{\partial P}{\partial r}=-G\frac{M_c}{r^2}\rho.
\label{eq:pressure}
\end{eqnarray}
where $G$ is the gravitational constant and 
$M_c$ is the mass of the core, which we take to be much larger than
the mass of the gaseous envelope. Equation (\ref{eq:pressure})
is not applicable if envelope is rapidly rotating (when the 
azimuthal velocity of gas is of the order of the local 
Keplerian velocity around the core), thus in the following 
discussion we assume envelope rotation to be slow.

Atmosphere is assumed to be heated from 
below by a source at the core surface with luminosity $L$. 
This energy can be transported by radiative diffusion or 
convection. We use the Schwarzschild criterion to determine
the convective stability of the envelope:
\begin{eqnarray}
\nabla<\nabla_{ad}\equiv\frac{\gamma-1}{\gamma},~~~
\nabla\equiv \frac{\partial\ln T}{\partial\ln P},
\label{eq:stab_cond}
\end{eqnarray}
where $\nabla$ is a temperature gradient, $\nabla_{ad}$ is its 
value under isentropic conditions. As usual, $\gamma$ is a 
adiabatic index of the gas; for monoatomic gas $\nabla_{ad}=2/5$, 
for diatomic (e.g. H$_2$) $\nabla_{ad}=2/7$. Usage of the  
Schwarzschild criterion implies that envelope
is chemically homogeneous and nonrotating --- a more 
general Ledoux criterion should be used in the presence of the 
molecular weight gradient (Kippenhahn \& Weigert 1990),  
and H\o iland criterion has to be employed if envelope
rotates rapidly (Tassoul 1978).

When atmosphere is convectively stable according to 
(\ref{eq:stab_cond}), energy released at the core surface 
has to be carried away radiatively. In this case 
one supplements (\ref{eq:pressure}) with the equation of 
radiation transfer. In the diffusion approximation, valid in 
the optically thick case, it reads
\begin{eqnarray}
\frac{16\sigma T^3}{3\kappa \rho}\frac{\partial T}{\partial r}=
-\frac{L}{4\pi r^2},
\label{eq:rad_trans}
\end{eqnarray}
where $\sigma$ is a Stefan-Boltzmann constant and $\kappa$ is the 
opacity. In the outer Solar System gas around the protoplanet 
can be so rarefied that the outer parts 
of the envelope are optically thin. This possibility is considered in 
more detail in \S \ref{subsect:opt_thin}. 

Whenever the stability criterion (\ref{eq:stab_cond}) is violated  
energy in the envelope is transported by convection. 
In this study we assume that convection is so effective 
that the temperature gradient in the convective parts of the 
envelope is equal to the adiabatic temperature gradient $\nabla_{ad}$.
 This is equivalent to supplementing (\ref{eq:pressure}) 
with adiabatic equation of state (isentropic gas)
\begin{eqnarray}
P=K\rho^\gamma,
\label{eq:polytrope} 
\end{eqnarray}
where $K$ is the adiabatic constant --- measure of the gas entropy. 
This approximation should be good enough in the dense regions  
of the envelope (very good in the interiors of present day 
giant planets). 
In Appendix \ref{ap:conv_efficiency} we determine under what circumstances 
this zero entropy gradient assumption is valid in the atmospheres of 
protoplanetary cores.

We suppose the luminosity $L$ to be derived from
the accretion of planetesimals and neglect the additional energy release
due to the radioactive heating and differentiation inside the core.
Thus we take
\begin{eqnarray}
L=G\frac{M_c\dot M}{R_c},
\label{eq:luminosity}
\end{eqnarray}
where $\dot M$ is a planetesimal accretion rate.
In Appendix \ref{ap:acc_rate} we briefly summarize three different 
regimes of planetesimal accretion  
important for the core growth, and calculate $\dot M$
and accretion timescale for each of them.
Equation (\ref{eq:luminosity}) assumes that
(a) planetesimal velocity at infinity relative to the core is not too large 
compared to the core's escape speed, and (b) planetesimals penetrate to 
the core surface without much resistance from the envelope
and release there all their kinetic energy. First 
assumption is quite reasonable during the buildup of the core
by planetesimal accretion; second should be valid for large 
planetesimals but small ones may be slowed down by the gas drag in 
the envelope as they make their way to the core surface. In the latter
case accretion energy release does not occur exactly at the core 
surface but is distributed throughout the envelope, meaning that  
luminosity depends on $r$. However,
even in the most unfavorable case of small planetesimals which are 
quasi-statically lowered from the top of the atmosphere to its bottom, 
luminosity is $(1-R_c/r)L$, where $R_c$ is the core radius, 
i.e. luminosity is not constant only
very near the core's surface (Pollack \etal 1996). Beyond several 
$R_c$, in the bulk of 
the envelope, we can still safely assume that $L$ is constant and
given by (\ref{eq:luminosity}). 

Use of the steady state equations (\ref{eq:pressure})
and (\ref{eq:rad_trans}) tacitly assumes that envelope can quickly adjust 
to the changes in the core mass $M_c$ and luminosity $L$ caused by the 
planetesimal accretion. In other words, these equations hold only 
provided that the dynamical and thermal timescales of the envelope are 
shorter than the core accretion timescale, and we demonstrate in 
Appendix \ref{ap:thermal_time} that this assumption is reasonable. 
We also show there that in the framework of quasi-stationary approximation 
energy release within the envelope due to gas accretion 
can be naturally neglected compared to the planetesimal accretion 
luminosity (\ref{eq:luminosity}).


\subsection{Important length scales.}
\label{sect:length_scales}

There are several characteristic length scales which are important
for the problem at hand. One of them is the
so-called Hill radius $r=R_H$
defined by
\begin{eqnarray}
R_H\equiv a\left(\frac{M_c}{M_\odot}\right)^{1/3}\approx 2\times 
10^{11}~\mbox{cm}~a_1\left(\frac{M_c}{M_\oplus}\right)^{1/3},
\label{eq:Hill}
\end{eqnarray}
Fluid elements at about $R_H$ from the protoplanetary core are 
equally affected by the core gravity and the tidal field 
of the central star. 
Within the Hill sphere gas dynamics is determined by the 
gravity of the core, while outside of it gas
is subject mostly to the stellar gravity.

Physical size of the core $R_c$ scales with $M_c$ in the same way 
as $R_H$ does, meaning that their ratio $p$ is a 
constant depending only on the physical density of the protoplanet
$\rho_c$ and the core location in the protoplanetary disk:
\begin{eqnarray}
p\equiv\frac{R_c}{R_H}=
\left(\frac{3}{4\pi}\frac{M_\odot}{\rho_c a^3}\right)^{1/3}\approx
5.2\times 10^{-3}~a_1^{-1}\rho_1^{-1/3},
\label{eq:p_def}
\end{eqnarray} 
where $\rho_1\equiv \rho/(1\mbox{ g cm}^{-3})$.
One can easily see that $R_c\ll R_H$.  

Bondi radius $R_B$  is defined as the 
distance from the protoplanet at which the thermal energy of the 
nebular gas is of the order of its gravitational
energy in the potential well of the core:
\begin{eqnarray} 
R_B\equiv G\frac{M_c}{c_0^2}\approx 4\times 
10^{10}~\mbox{cm}~a_1^{1/2}\frac{M_c}{M_\oplus}.
\label{eq:Bondi}
\end{eqnarray}
Outside Bondi sphere ($r\gtrsim R_B$) gravity of the core 
is too weak to strongly affect the gas; consequently, gas 
pressure is almost equal to its nebular value $P_0$. 
Inside Bondi sphere pressure is significantly perturbed 
by the protoplanetary gravity. 

\begin{figure}
\plotone{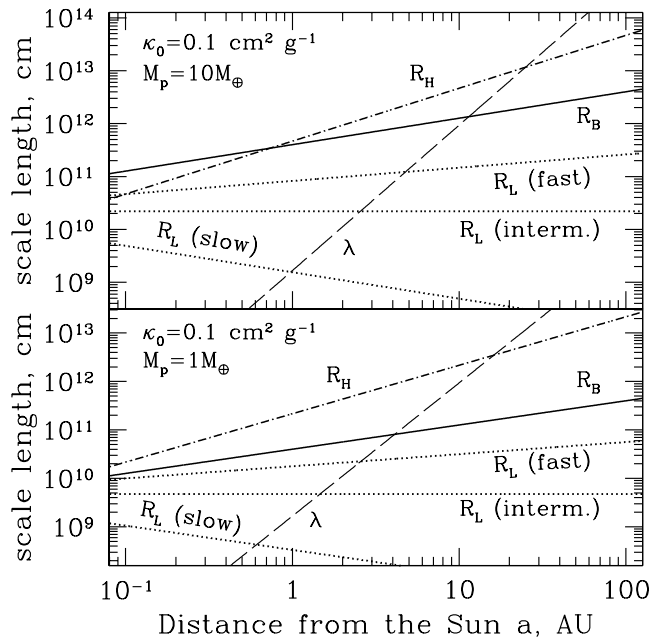}
\caption{Different length scales important for atmospheric 
structure as a function of semi-major axis $a$ for
$\kappa_0=0.1$ cm$^2$ g$^{-1}$ and two values 
of core mass: $M_c=10$ M$_\oplus~ (top)$ and  $M_c=1$ M$_\oplus~ 
(bottom)$. Solid, dashed, and dot-dashed lines represent 
$R_B$, $\lambda$, and $R_H$ correspondingly. Dotted lines
correspond to luminosity radius $R_L$ evaluated for three
different planetesimal accretion regimes --- fast, intermediate, 
and slow (see Appendix \ref{ap:acc_rate}) --- labeled on the 
plot. The meaning of other curves is clear from the corresponding
labels.
\label{fig:scales}}
\end{figure}

Opacity of the gas sets another important scale.
We consider a rather general opacity law by assuming
throughout this work that 
\begin{eqnarray}
\kappa=\kappa_0\left(P/P_0\right)^{\alpha}
\left(T/T_0\right)^{\beta}.
\label{eq:opacity}
\end{eqnarray} 
Previous studies (Mizuno 1980; Stevenson 1982) have always assumed 
dust opacity in the outer layers of the envelope assuming it 
to be constant. At the same time, opacity due to small interstellar 
dust grains (smaller than the typical wavelength of 
the local blackbody radiation) is thought to behave as $\kappa\propto T^\beta$ 
(i.e. $\alpha=0$) with $\beta\approx 1-2$ (Draine 2003) in the range 
of temperatures typical for protoplanetary disks  
(although this is a statement depending on the size distribution 
and composition of the dust, which in the protoplanetary 
disks can be different 
from those in the ISM). Our opacity prescription (\ref{eq:opacity}) 
accounts for this possibility and is more general compared to the 
previous treatments. Opacity in the inner layers of the envelope (likely dominated 
by the molecular opacity of H$_2$ and H$_2$O for which
one has to use opacity tables) might not be so 
important for the envelope structure or mass --- atmosphere 
can be convective there, see \S \ref{subsect:comparison}. 
Opacity effects are quantified 
by introducing a mean absorption length of
photons in the nebular gas 
\begin{eqnarray}
\lambda\equiv (\kappa_0\rho_0)^{-1}\approx 1.7 \times 
10^9~\mbox{cm}~a_1^{11/4}
\left(\frac{0.1~\mbox{cm$^2$ g$^{-1}$}}{\kappa_0}\right).
\label{eq:lambda}
\end{eqnarray} 
The exact value of $\kappa_0$
is highly uncertain because the amount of dust and its size
distribution in protoplanetary disks are poorly constrained. 
Thus, we treat $\kappa_0$ as a parameter and for simplicity 
take it to be constant throughout the nebula, independent of $a$. 
Outer parts of the protoplanetary nebula are optically thin,
meaning that $\lambda$ is larger than the vertical disk scale  
height $h\equiv c_0/\Omega$. 

Finally, luminosity of the core sets one more length scale
\begin{eqnarray}
R_L\equiv \left(\frac{L}{16\pi \sigma T_0^4}\right)^{1/2},
\label{eq:R_L}
\end{eqnarray} 
which is a size of object radiating luminosity $L$ at
an effective temperature $T_0$. In Figure \ref{fig:scales} 
we plot $R_L$ together with other length scales as a function 
of $a$ for two values of $\kappa_0$ and for different accretion
regimes (see Appendix \ref{ap:acc_rate}). 

Nebula can be considered  homogeneous on the scale of 
$R_B$ only if
$R_B$ is small compared to the disk scaleheight $h$. 
This condition puts the following 
constraint on the mass of the core:
\begin{eqnarray}
M_c\ll M_1\equiv \frac{c_0^3}{G\Omega}\approx 12~M_\oplus~
a_1^{3/4}.
\label{eq:M_1}
\end{eqnarray}
Whenever this condition is 
fulfilled the Bondi radius is also smaller than $R_H$, 
while $R_H$ is smaller than $h$, see equation (\ref{eq:r_b}). 
For $M_c\gg M_1$ Bondi radius lies outside $R_H$ 
and nebular gas in the Hill sphere is very inhomogeneous 
($R_H\gg h$) and 
strongly perturbed by the protoplanetary gravity. At $10$ 
AU (\ref{eq:M_1}) constrains 
$M_c$ to be less than $\approx 70 M_\oplus$, while
at $30$ AU $M_1\approx 150 M_\oplus$. Note that (\ref{eq:M_1}) 
is also an approximate condition for the absence of
strong spatial gradients in the nebula caused by 
the dissipation of the core-induced density waves 
(Lin \& Papaloizou 1993; Rafikov 2002). As a 
result, one does not have to worry about the gap formation.

\begin{figure}
\plotone{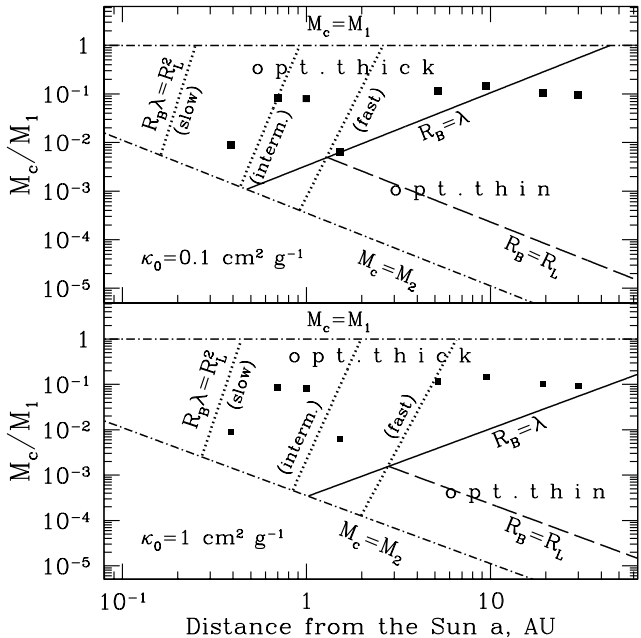}
\caption{Relationships between different length scales 
as a function of semi-major axis $a$ in the nebula for 
two values of gas opacity (assumed constant throughout 
the nebula): $\kappa_0=0.1$ cm$^2$ g$^{-1}$ ({\it top}) and
$\kappa_0=1$ cm$^2$ g$^{-1}$ ({\it bottom}). Only the case 
$M_2<M_c<M_1$ is considered (dot-dashed boundaries).
Dotted lines correspond to condition $R_B\lambda=R_L^2$
for three different planetesimal accretion regimes labeled 
on the curves. Solid line represents $R_B=\lambda$ separating
optically thick and optically thin regimes (see \S 
\ref{subsect:pure_dif} and \ref{subsect:opt_thin}), while
dashed line is for $R_B=R_L$. Squares mark positions of 8 
major planets of the Solar System on this diagram; for 
Jupiter and Saturn we assume cores of $5$ M$_\oplus$ and 
$10$ M$_\oplus$.
\label{fig:regions}}
\end{figure}

With the use of (\ref{eq:M_1}) one can rewrite (\ref{eq:Bondi}) in the
following form:
\begin{eqnarray}
R_B=R_H\left(\frac{M_c}{M_1}\right)^{2/3}=R_c
\left(\frac{M_c}{M_2}\right)^{2/3},
\label{eq:r_b}
\end{eqnarray}
where
\begin{eqnarray}
M_2\equiv p^{3/2}M_1
\approx 4.5\times 10^{-3}~M_\oplus~a_1^{-3/4}\rho_1^{-1/2}
\label{eq:R_B_rat}
\end{eqnarray}
is another fiducial mass scale. If the core mass $M_c$ is smaller than $M_2$ 
then $R_B\lesssim R_c$, meaning that the protoplanet is too small to induce
appreciable pressure perturbations in the surrounding gas even at its 
surface and thus has no atmosphere associated with it. 
We will study only the cores satisfying
\begin{eqnarray}
M_2\ll M_c\ll M_1
\label{eq:cond}
\end{eqnarray} 
since they are
massive enough to possess atmospheres and small enough for the surrounding 
gas to be thought of as roughly homogeneous on the scale of $R_B$, i.e.
$R_c\ll R_B \ll R_H\ll h$. 
One can see that the mass range in which (\ref{eq:cond}) 
is valid is rather large and spans $4-5$ orders of magnitude depending 
on $a$ (e.g. from $\sim 0.1$ Lunar mass to $\sim M_J$ at $10$ AU).
In Figure \ref{fig:regions} we demonstrate how the different length 
scales are related to each other as a function of $a$ and $M_c/M_1$,
constrained by (\ref{eq:cond}).


\subsection{Boundary conditions.}
\label{sect:boundary}

Boundary conditions to the equations of \S \ref{sect:struct} specify
that gas pressure, temperature, and density should reduce
to their nebular values $P_0$, $T_0$, and $\rho_0$ 
at some distance $R_{out}$ from the core:
\begin{eqnarray}
T(R_{out})=T_0,~~~P(R_{out})=P_0,~~~\rho(R_{out})=\rho_0.
\label{eq:bound}
\end{eqnarray}
The problem then is to determine the value of $R_{out}$.

The most simplistic way of evaluating $R_{out}$
would be to completely neglect the fact that the nebula 
surrounding the core is in the differential motion caused 
by the gravity of the central star. If one does this and 
assumes gas to be static everywhere, then $R_{out}$ can be 
set equal to infinity. Presence of the shearing motion in the 
nebula drastically changes the dynamical and thermal behavior 
of the gas in the core's vicinity. First, gas inside the Hill sphere 
moves in a rather complicated way as demonstrated by the  the 
hydrodynamical simulations
of planet-disk interaction (e.g. D'Angelo \etal 2002, 2003); but we 
can hope that this will not violate our assumption of nebula homogeneity 
in the core vicinity if $R_B\lesssim R_H$ (or $M_c\lesssim M_1$). 

Second, the flow of the nebular gas within the Hill sphere
can effectively ``cool'' the outer part of the envelope by advecting 
the gas heated by the core's accretional energy release away from the 
core and bringing in fresh gas having entropy equal to the nebular 
entropy. This process should determine the value of $R_{out}$ at which 
temperature drops to $T_0$. Apparently, $R_{out}\gtrsim R_B$ since gas
within Bondi sphere is confined by the core's gravity
and is not being refreshed. 
At the same time, the distance from the core center at 
which the boundary condition for the pressure  
is imposed is determined by a different process ---
dominance of the core gravity over the thermal pressure 
in the nebula (and, possibly, over the centrifugal forces 
induced by the fluid motion around the gravitationally 
bound part of the envelope which we neglect in this study). 
Apparently, pressure may converge to $P_0$ only beyond $R_B$;  
thus, we again arrive at the restriction $R_{out}\gtrsim R_B$
but now from the point of view of pressure balance.

A proper estimate of distance $R_{out}$ at which the thermal 
and pressure 
boundary conditions are imposed must include inhomogeneous
heating within $R_{out}$, geometry of the flow 
near the core, vertical structure of the nebula, possibility 
of formation of rotationally supported disk around the 
core, and so on. This is beyond the scope of this study. 
Fortunately, it turns out (see \S \ref{subsect:assumptions}) 
that the exact value of $R_{out}$ 
is not important for the envelope structure as long as 
$R_{out}\gg R_B$ (we checked this to be true); in our
calculations we simply set $R_{out}=20 R_B$.


\section{Atmospheres with outer radiative zone.}
\label{sect:solutions}
 

Relationships between the various length scales defined in
\S \ref{sect:length_scales} are quite different in the inner 
and outer parts of the nebula. As Figure \ref{fig:regions}
demonstrates, for protoplanetary cores with 
$M_c\approx (1-10) M_\oplus$ accreting planetesimals 
in the fast regime at $a\gtrsim (2-5)$ AU, one of the 
two sets of inequalities 
\begin{eqnarray}
&& \lambda\ll R_B,~~~R_L^2\ll \lambda R_B,~~~~~\mbox{or} 
\label{eq:inequality2}
\\ 
&& R_L\ll R_B\ll \lambda,
\label{eq:inequality1}
\end{eqnarray}
is typically fulfilled depending on $a$ and $\kappa_0$. This is 
mainly because the planetesimal accretion rate $\dot M$ and nebular 
gas density $\rho_0$ decrease with $a$ reducing $R_L$ and 
increasing $\lambda$. For 
smaller cores or slower accretion regime these conditions can hold 
even closer to the Sun. We demonstrate later on that when either  
(\ref{eq:inequality1}) or (\ref{eq:inequality1}) is valid, 
there exists an almost isothermal radiative layer in the outer 
part of the envelope  
which decouples its interior from the nebular gas. Although the 
structure of this layer is slightly different in two 
cases, properties of the inner envelope will
be shown to be universal.


\subsection{Low luminosity, optically thick 
($\lambda\lesssim R_B$ and $R_L^2\ll\lambda r_B$).}
\label{subsect:pure_dif}

Whenever the photon mean-free path $\lambda$ is shorter 
than the Bondi radius $R_B$, envelope is optically thick to 
the escaping radiation. In this case its 
structure is completely determined by equations (\ref{eq:pressure}) 
and (\ref{eq:rad_trans}). Integrating them together with 
(\ref{eq:opacity}) and using boundary condition $P=P_0$ when 
$T=T_0$ (at $r=R_{out}$, see [\ref{eq:bound}]) one finds that
\begin{eqnarray}
&&\left(\frac{P}{P_0}\right)^{1+\alpha}-1
=\frac{4\nabla_0}{3}
\frac{R_B \lambda}{R_L^2}\left[
\left(\frac{T}{T_0}\right)^{4-\beta}-1\right],\nonumber\\
&& \nabla_0\equiv\frac{1+\alpha}{4-\beta}
\label{eq:purely_rad_sol}
\end{eqnarray}
Constant $\nabla_0$ 
is positive whenever $T$ increases with depth; we assume this
to be the case which requires $\beta<4$ since normally opacity 
does not decrease with increasing density (i.e. $\alpha\ge 0$). 
The meaning of $\nabla_0$ will become clear later on 
(see \S \ref{subsect:convection}).
The coefficient in the right hand side of
(\ref{eq:purely_rad_sol}) is very large because of 
(\ref{eq:inequality2}), so that a large change 
of pressure results in only a small perturbation of 
temperature.

To make further progress we need to substitute
(\ref{eq:purely_rad_sol}) into either 
(\ref{eq:pressure}) or (\ref{eq:rad_trans}) and
find the $P$ and $T$ dependencies on $r$. This 
can be done numerically for arbitrary
$\alpha$ and $\beta$. Here we look at the asymptotic
behavior of the envelope properties in two limiting cases: in the
outer atmosphere where 
$T-T_0\lesssim T_0$, and deep in the envelope
where $T\gg T_0$. 

\begin{figure}
\plotone{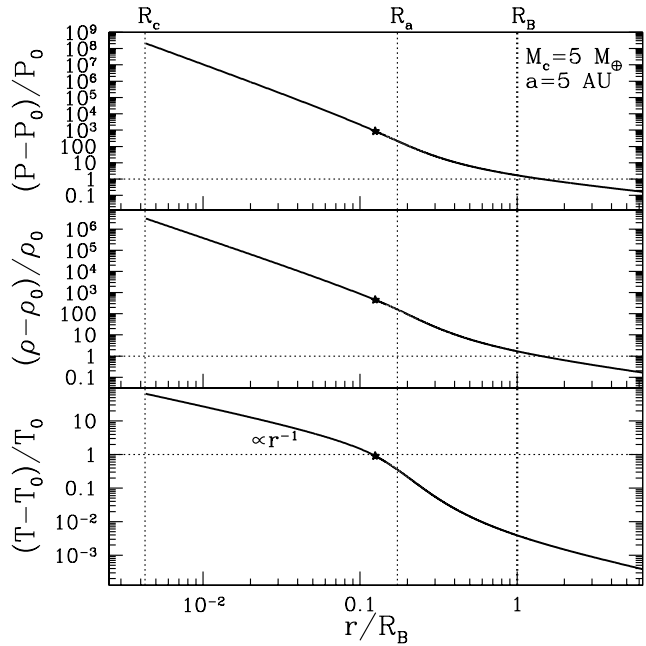}
\caption{Atmospheric structure around the core with 
$M_c=5~M_\oplus$ at $5$ AU. Relative deviations of $P$, 
$\rho$, and $T$ from their nebular values calculated numerically 
are shown 
($P_0=3.2\times 10^{-7}$ bar, $\rho_0=7\times 10^{-11}$ 
g cm$^{-3}$ and $T_0=130$ K in the surrounding nebula). 
Calculation is done for $\alpha=0,~\beta=1,~\gamma=7/5$
and $\kappa=0.1$ cm$^2$ g$^{-1}$.
In this particular case $\lambda<R_B$ and 
$R_B\lambda/R_L^2\approx 300$, i.e envelope is optically 
thick everywhere and has an outer radiative zone,
see \S \ref{subsect:pure_dif}. Stars mark the position 
of the outer edge of the inner convective region. 
\label{fig:rad_str_5_5}}
\end{figure}

In the first case envelope is essentially isothermal; solving
(\ref{eq:pressure}) under this assumption one obtains 
\begin{eqnarray}
P/P_0=\rho/\rho_0=\exp\left(\frac{R_B}{r}-\frac{R_B}{R_{out}}\right).
\label{eq:exp_structure}
\end{eqnarray}
Temperature profile can then be 
simply obtained from (\ref{eq:purely_rad_sol}) as
\begin{eqnarray}
&& \frac{T-T_0}{T_0}\approx \frac{3}{4(1+\alpha)}
\frac{R_L^2}{R_B\lambda}\nonumber\\
&& \times\left\{\exp\left[(1+\alpha)
\left(\frac{R_B}{r}-\frac{R_B}{R_{out}}\right)
\right]-1\right\},
\label{eq:t_small}
\end{eqnarray}
where boundary conditions (\ref{eq:bound}) were again used. 
One can see from (\ref{eq:t_small}) that
temperature perturbation at $r\sim R_B$ is small because 
$R_B\lambda/R_L^2\gg 1$; at the same time gas density 
and pressure within Bondi sphere 
grow exponentially, see (\ref{eq:exp_structure}). Gas 
temperature rises to several times $T_0$ only when the 
distance to the core center becomes as small as
\begin{eqnarray}
R_a\equiv R_B\frac{1+\alpha}{\ln (R_B \lambda /R_L^2)}.
\label{eq:R_a}
\end{eqnarray}
Apparently, $R_a\ll R_B$ although 
one should keep in mind that $R_a$ depends on $R_B\lambda/R_L^2$ 
only logarithmically. 
Pressure $P_a$ and density $\rho_a$ at this depth are
\begin{eqnarray}
P_a/P_0\approx \rho_a/\rho_0\approx 
(R_B \lambda/R_L^2)^{1/(1+\alpha)}\gg 1.
\label{eq:P_a}
\end{eqnarray}
Existence of the outer radiative layer, in which temperature is 
almost constant while density dramatically increases was first found 
numerically by Harris (1978) under the conditions typical in 
the region of giant planets. Note that the presence of the outer 
isothermal layer does not necessarily require 
gas to be optically thin, cf. Mizuno \etal (1978).

In a second limiting case, interior to $R_a$, we use $T\gg T_0$ 
to integrate (\ref{eq:rad_trans}) with (\ref{eq:purely_rad_sol}). 
Using $T\approx \xi T_0 ~(\xi\sim 1)$ at $r=R_a$ as an 
approximate boundary condition at the transition point
one finds that
\begin{eqnarray}
&& \frac{T}{T_0}\approx \xi+\nabla_0
\left(\frac{R_B}{r}-\frac{R_B}{R_a}\right),
\label{eq:rad_zero_t}\\
&& \frac{P}{P_0}\approx 
\left(\frac{4\nabla_0}{3}\frac{\lambda R_B}{R_L^2}
\right)^{1/(1+\alpha)}
\left[\xi+\nabla_0
\left(\frac{R_B}{r}-\frac{R_B}{R_a}\right)\right]^{1/\nabla_0}.
\label{eq:rad_zero_p}
\end{eqnarray}
In the case of constant opacity ($\alpha=\beta=0,~\nabla_0=1/4$)
this solution reduces to the radiative
zero solution found by Stevenson (1982) for $r\ll R_a$. 
Figure \ref{fig:rad_str_5_5} shows the internal structure of 
the atmosphere around $5$ M$_\oplus$ core at $5$ AU (for 
which [\ref{eq:inequality2}] is valid) calculated using (\ref{eq:pressure}) 
and (\ref{eq:purely_rad_sol}). Opacity with $\alpha=0,~\beta=1$ 
and $\kappa_0=0.1$ cm$^2$ g$^{-1}$ as well as $\gamma=7/5$ 
are assumed in calculation. This specific scaling of
$\kappa$ with $P$ and $T$ makes the inner part of the 
envelope convective below $R_a$ (see \S \ref{subsect:convection}),
but this hardly changes the overall picture of the 
atmospheric structure that we described in this section.


\subsection{Low  luminosity, optically thin
($R_L\ll R_B\lesssim\lambda$).}
\label{subsect:opt_thin}

Whenever  $\lambda\gtrsim R_B$  there is  
an optically thin region around the protoplanet in which 
the photon mean free path is larger than the 
length scale over which physical variables such as pressure
and temperature experience significant changes. 
Deep in the envelope (below the photosphere) 
optical depth increases and atmosphere becomes optically thick. 
Gas is optically thick also far from the core since the density 
and temperature scale height is $\sim r$ there and the 
optical depth to radiation escaping from the core becomes 
larger than unity at $r\gtrsim \lambda$. 
Of course, this outer optically thick region can only exist if 
$\lambda\lesssim h$ and nebula is homogeneous on scales $\sim\lambda$, 
which is not the case in the outer parts of protoplanetary 
disks (but this turns out not to be important). 
Thus, the optically thin zone around the core 
in the infinite medium should be sandwiched between the 
inner and outer optically thick regions. 

Temperature structure in the optically thick parts is determined 
by equation (\ref{eq:rad_trans}). In the optically thin region
we have $T^4\approx T_0^4+L/(16\pi\sigma r^2)$, where the additional
factor of $4$ in the denominator of the second term comes from
the anisotropy of the core radiation. Similar behavior of $T$ in the 
optically thin region was found by Hayashi \etal (1979). Strictly 
speaking, this expression is accurate only far from the 
photosphere and we should expect it to reproduce photospheric 
temperature only approximately (which will not affect our results in
any significant way). Thus, we express envelope temperature profile 
above the photosphere as
\begin{eqnarray}
T^4(r)\approx T_0^4+\frac{L}{16\pi\sigma r^2}
+\frac{3L}{16\pi\sigma}\int\limits_r^\infty\frac{\kappa\rho(r^\prime)dr^\prime}
{r^{\prime 2}}
\label{eq:t_profile}
\end{eqnarray}
The second term on the right hand side is relevant for the  
optically thin region while the third becomes important outside 
of $r\approx \lambda$ where photons couple to the nebular gas and 
start to diffuse. 

Gas pressure is close to $P_0$ for $r\gtrsim R_B$. Assuming that 
$T-T_0\ll T_0$ everywhere in this region
we have $\rho\approx \rho_0$ and $\kappa\approx\kappa_0$.
Then it follows from (\ref{eq:t_profile}) that
\begin{eqnarray}
T(r)\approx T_0\left[1+\frac{R_L^2}{r^2}
+3\frac{R_L^2}{\lambda r}\right]^{1/4}.
\label{eq:t_approx}
\end{eqnarray}  
It is very important that the temperature perturbation at the Bondi sphere 
$T(R_B)-T_0\sim T_0 (R_L/R_B)^2$ is negligible whenever (\ref{eq:inequality1})
is fulfilled (similar to the optically thick case considered in 
\S \ref{subsect:pure_dif}) --- this verifies our assumption of $T\approx T_0$ 
for $r\gtrsim R_B$. This result is independent 
of whether $\lambda$ is larger or smaller than $h$ or $R_H$, what 
is crucial is that $R_L\ll R_B\ll \lambda$ (which makes third term
in [\ref{eq:t_approx}] negligible compared to the second at $r\sim R_B$). 
Temperature deviation
$T-T_0$ is dominated by the radiative diffusion for 
$r\gtrsim \lambda$ (third term in brackets) and by the 
optically thin radiation transfer for $r\lesssim \lambda$.

\begin{figure}
\plotone{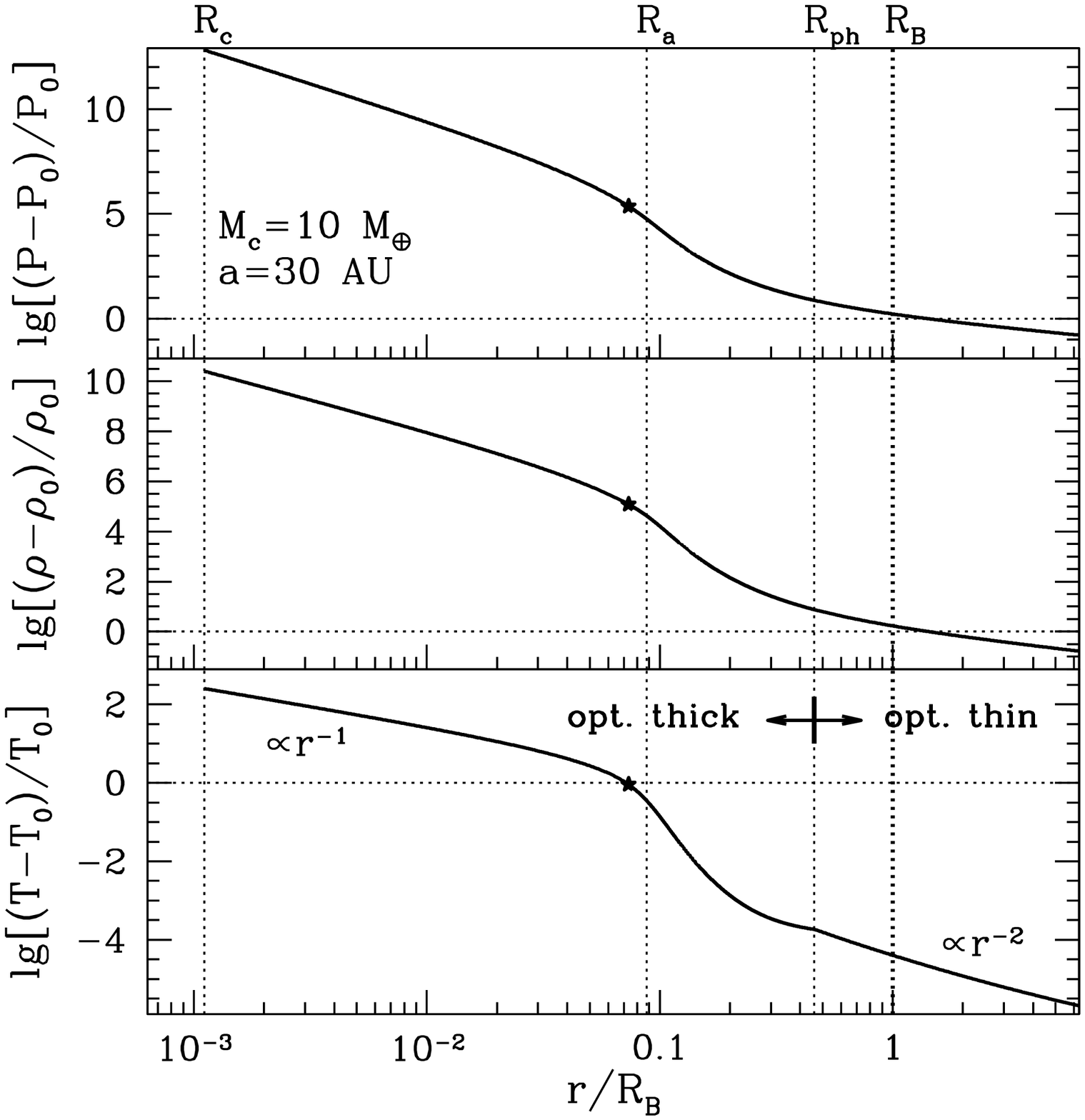}
\caption{Same as Figure \ref{fig:rad_str_5_5} but for 
$M_c=10~M_\oplus$ at $30$ AU. In
this particular case $P_0=10^{-9}$ bar, 
$\rho_0=5\times 10^{-13}$ g cm$^{-3}$ and $T_0=55$ K.  Also 
$\lambda > R_B$ and 
$R_B\lambda/R_L^2\approx 10^5$, i.e envelope has an outer 
optically thin region and possesses an outer radiative zone,
see \S \ref{subsect:opt_thin}. The rest of notation is as in 
Figure \ref{fig:rad_str_5_5}.
\label{fig:rad_str_10_30}}
\end{figure}

Inside the Bondi sphere gas is initially still optically thin 
and its temperature is  
determined by (\ref{eq:t_approx}) meaning that $T\approx T_0$. 
At the same time,  
pressure and density in this  essentially isothermal region 
increase exponentially 
with depth in accordance with (\ref{eq:exp_structure}). 
As a result, local photon mean free path rapidly decreases
and envelope finally becomes optically thick to the 
escaping radiation. Photosphere is located at the distance $R_{ph}$
from the core center where the photon mean free path becomes 
comparable to the typical length scale of density variation.  
Using (\ref{eq:exp_structure}) we estimate 
that the density scale height is $\partial r/\partial\ln\rho = 
r^2/R_B$, which becomes comparable to $(\kappa\rho)^{-1}$ at
\begin{eqnarray}
R_{ph}\approx R_B\frac{1+\alpha}{\ln\left(\lambda/R_B\right)},
\label{eq:r_ph}
\end{eqnarray}
whenever $R_B\ll\lambda$. We find from (\ref{eq:t_approx}) that
gas temperature at the photosphere $T_{ph}\equiv T(R_{ph})$ 
is offset from $T_0$ by
\begin{eqnarray}
\frac{T_{ph}-T_0}{T_0}\approx\frac{R_L^2}{4R_{ph}^2}\approx
\left[\frac{\ln\left(\lambda/R_B\right)}{2(1+\alpha)}
\frac{R_L}{R_B}\right]^2\ll 1,
\label{eq:t_phot}
\end{eqnarray}
while the photospheric pressure $P_{ph}\equiv P(R_{ph})$ is
\begin{eqnarray}
P_{ph}/P_0\approx (\lambda/R_B)^{1/(1+\alpha)}\gg 1.
\label{eq:p_phot}
\end{eqnarray}
We see that $T_{ph}$ is still only slightly 
different from $T_0$, while the photospheric pressure is much 
higher than $P_0$ if (\ref{eq:inequality1}) holds.

Below the photosphere, 
we again resort to equations (\ref{eq:pressure}) 
and (\ref{eq:rad_trans}). Similar to 
(\ref{eq:purely_rad_sol}) one finds for boundary conditions 
$P=P_{ph}$ at $T=T_{ph}$ that  
\begin{eqnarray}
&& \left(\frac{P}{P_0}\right)^{1+\alpha}-
\left(\frac{P_{ph}}{P_0}\right)^{1+\alpha}\nonumber\\
&& =\frac{4\nabla_0}{3}
\frac{R_B \lambda}{R_L^2}\left[
\left(\frac{T}{T_0}\right)^{4-\beta}-
\left(\frac{T_{ph}}{T_0}\right)^{4-\beta}\right].
\label{eq:rad_sol1}
\end{eqnarray} 
When $|T-T_0|/T_0\lesssim 1$ we can still use (\ref{eq:exp_structure})
for the density profile; then the temperature profile 
for $r\lesssim R_{ph}$ derived from 
(\ref{eq:rad_sol1}) becomes 
\begin{eqnarray}
&& \frac{T-T_{ph}}{T_0}\approx \frac{3}{4(1+\alpha)}
\frac{R_L^2}{R_B\lambda}\nonumber\\
&& \times\left\{\exp\left[(1+\alpha)
\frac{R_B}{r}
\right]-(P_{ph}/P_0)^{1+\alpha}\right\},
\label{eq:t_small_thin}
\end{eqnarray}
where we took into account that $R_B/R_{out}\lesssim 1$.
Keeping in mind that according to (\ref{eq:inequality1})
and (\ref{eq:p_phot}) 
$(P_{ph}/P_0)^{1+\alpha}\ll R_B\lambda/R_L^2$,
 one easily finds that the distance
$R_a$ at which gas temperature 
starts to appreciably deviate from $T_0$ is still given by
(\ref{eq:R_a}). Thus, under conditions 
(\ref{eq:inequality1}) $R_a$ is still
located deep inside the envelope, and always below the 
photosphere, compare with (\ref{eq:r_ph}). 

It is also easy to deduce from (\ref{eq:rad_sol1}) 
that for $r\lesssim R_a$ temperature and pressure behavior
in the inner envelope are still described by equations 
(\ref{eq:rad_zero_t}) and (\ref{eq:rad_zero_p}). Thus,
despite the differences in the structure of the outer
atmosphere ($r\gtrsim R_a$) in the optically thin 
($R_L\ll R_B\ll \lambda$) and optically thick 
($\lambda\ll R_B,~R_L^2\ll\lambda r_B$) cases, the structure 
of the inner atmosphere is the same. In Figure 
\ref{fig:rad_str_10_30} properties of the envelope
around $10$ M$_\oplus$ core at $30$ AU (for which 
[\ref{eq:inequality1}] is valid) are exhibited. Temperature 
profile in the optically thin region was calculated using 
(\ref{eq:t_profile}), and all parameters (opacity, value 
of $\gamma$) are the same as those used for 
Figure \ref{fig:rad_str_5_5},
see \S \ref{subsect:pure_dif}.


\subsection{Convective stability.}
\label{subsect:convection}

In \S \ref{subsect:pure_dif}-\ref{subsect:opt_thin} we have 
assumed that energy transfer from the bottom 
to the top of the atmosphere occurs by radiative transport. 
Here we explicitly determine under which circumstances 
this is the case, and also calculate envelope structure 
in convectively unstable regions.

To check our solutions obtained in 
previous sections for convective stability  we calculate $\nabla$ and 
use the Schwarzschild criterion (\ref{eq:stab_cond}). It 
turns out that the outer part of the 
envelope above $R_a$ is stable whenever (\ref{eq:inequality2}) or 
(\ref{eq:inequality1}) is fulfilled.
Indeed, in the optically thick case (see \S \ref{subsect:pure_dif})
we find using (\ref{eq:pressure}), (\ref{eq:rad_trans}), 
and (\ref{eq:purely_rad_sol}) that 
\begin{eqnarray}
&& \nabla(T)=\frac{3}{4}\frac{R_L^2}{R_B\lambda}
\left(\frac{T}{T_0}\right)^{\beta-4}
\left(\frac{P}{P_0}\right)^{1+\alpha} \nonumber\\
&& =
\nabla_0\left\{1-\left(\frac{T_0}{T}\right)^{4-\beta}
\left[1-\frac{3}{4\nabla_0}\frac{R_L^2}{R_B\lambda}
\right]\right\}.
\label{eq:nabla_in}
\end{eqnarray}
Because of the limitations imposed by (\ref{eq:inequality2}) this 
reduces to
\begin{eqnarray}
\nabla\approx \nabla_0
\left[1-\left(\frac{T_0}{T}\right)^{4-\beta}
\right],
\label{eq:nabla_in1}
\end{eqnarray}
[relative corrections to this expression are at the most 
$O(R_L^2/R_B\lambda)\ll 1$]. Thus, outside of $R_a$, where
$T\approx T_0$, temperature gradient is small and
atmosphere is convectively stable. We also see from (\ref{eq:nabla_in1})
that $\nabla_0$ has a meaning of 
temperature gradient deep inside the radiative envelope where $T\gg T_0$.

In the optically thin case (see \S \ref{subsect:opt_thin})
one finds similar situation. Temperature above the photosphere
is almost equal to $T_0$ (see equations 
[\ref{eq:t_approx}] and [\ref{eq:t_phot}]) while pressure 
profile is given by (\ref{eq:exp_structure}). 
Then one obtains that
\begin{eqnarray}
\nabla\approx \frac{1}{4}\left(2\frac{R_L^2}{R_B r}
+3\frac{R_L^2}{R_B\lambda}\right)\ll 1,
\label{eq:nabla_out}
\end{eqnarray}
i.e. atmosphere is convectively stable above the photosphere.
Below the photosphere, for $r \lesssim R_{ph}$, 
temperature is related to pressure by (\ref{eq:rad_sol1})
which results in 
\begin{eqnarray}
&& \nabla(T)=\nabla_0\left\{1-\left(\frac{T_{ph}}{T}\right)^{4-\beta}
\right. \nonumber\\
&& \left. \times\left[1-\frac{3}{4\nabla_0}\frac{R_L^2}{R_B\lambda}
\left(\frac{P_{ph}}{P_0}\right)^{1+\alpha}
\left(\frac{T_0}{T_{ph}}\right)^{4-\beta}
\right]\right\}.
\label{eq:nabla_in_thin}
\end{eqnarray}
Since according to (\ref{eq:t_phot}) \& (\ref{eq:p_phot}) 
$T_{ph}\approx T_0$ and $(P_{ph}/P_0)^{1+\alpha}\approx 
\lambda/R_B$, 
one finds that $\nabla$ below $R_{ph}$ is
still given by (\ref{eq:nabla_in1}) with the 
relative accuracy $O(R_L^2/R_B^2)\ll 1$. Consequently, 
in the optically thin case (\ref{eq:inequality1}) atmosphere 
is convectively stable everywhere above $R_a$, analogous to
the optically thick case. Thus, in both cases considered in
\S \ref{subsect:pure_dif} and \ref{subsect:opt_thin} there is
an {\it outer radiative (convectively stable) region} in the  
atmosphere, which is almost isothermal. 

Deep in the envelope, below $R_a$, radiative temperature 
gradient is given by (\ref{eq:nabla_in1}) and steadily increases 
with depth (since $T$ monotonically goes up); deep inside the 
envelope $\nabla$ becomes equal to $\nabla_0$. Thus, whenever
\begin{eqnarray}
\nabla_0=\frac{1+\alpha}{4-\beta}<\nabla_{ad}=\frac{\gamma-1}{\gamma}
\label{eq:cond_rad}
\end{eqnarray}
the whole envelope is convectively stable and energy
is transferred from the bottom by radiative diffusion.
If opacity is independent of $P$ and $T$, i.e. 
$\alpha=\beta=0$, then $\nabla_0=1/4$ (Stevenson 1982) and  
envelope is convectively stable provided that 
it is composed of monoatomic or diatomic gas. 

In  the opposite case, when  
$\nabla_0>\nabla_{ad}$ envelope becomes
convectively unstable at some point. From (\ref{eq:nabla_in1}) 
we find that this happens when gas temperature reaches 
\begin{eqnarray}
T_{conv}\approx T_0\left(1-\frac{\nabla_{ad}}{\nabla_0}
\right)^{-1/(4-\beta)}
\label{eq:T_conv}
\end{eqnarray}
This critical temperature above which convection sets in 
is clearly not very different from $T_0$. Thus, $T_{conv}$
is achieved at $\approx R_a$ from the core center and 
we conclude that atmosphere becomes convective below $\approx R_a$. 
For example, when the dominant source of opacity in the outer 
atmosphere is dust with $\beta=1$, one should 
expect envelope to be convectively unstable if H$_2$ 
is its major constituent. In this case convection sets in 
at $T_{conv}=7^{1/3}T_0\approx 1.9 T_0$. The edge of the 
convection zone as shown in Figures \ref{fig:rad_str_5_5} 
and \ref{fig:rad_str_10_30} agrees with this estimate.

If envelope is convectively unstable below $R_a$, 
equation (\ref{eq:rad_trans}) cannot be used there. Instead, 
one has to utilize (\ref{eq:polytrope}) to relate 
pressure and density in (\ref{eq:pressure}).
Value of $K$ in (\ref{eq:polytrope}) is set by the conditions 
at the edge of the convection zone, i.e at $T=T_{conv}$. In 
the optically thin atmospheres (see 
\S \ref{subsect:opt_thin}) pressure at this point is set 
by (\ref{eq:rad_sol1}) to be 
\begin{eqnarray}
&& P_{conv}\equiv P_0
\left\{\left(\frac{P_{ph}}{P_0}\right)^{1+\alpha}\right.\nonumber\\
&& \left.+\frac{4\nabla_0}{3}
\frac{R_B \lambda}{R_L^2}\left[
\frac{1}{1-\nabla_{ad}/\nabla_0}-
\left(\frac{T_{ph}}{T_0}\right)^{4-\beta}\right]
\right\}^{1/(1+\alpha)}\label{eq:P_rough}\nonumber \\
&& \approx P_0\left[
\frac{4}{3}\frac{\nabla_{ad}\nabla_0}{\nabla_0-\nabla_{ad}}
\frac{R_B \lambda}{R_L^2}
\right]^{1/(1+\alpha)}.
\label{eq:P_conv}
\end{eqnarray}
The last line in (\ref{eq:P_conv}) follows from (\ref{eq:p_phot}) 
and $T_{ph}\approx T_0$.  Apparently, $P_{conv}\sim P_a\gg P_0$. 
In the optically thick atmospheres pressure is given by 
(\ref{eq:purely_rad_sol}), 
which can be obtained from  (\ref{eq:rad_sol1}) by setting
$P_{ph}=P_0$ and $T_{ph}=T_0$. Making the same substitutions in 
equation (\ref{eq:P_rough}) we find that $P_{conv}$ is still given 
by (\ref{eq:P_conv}), despite the different structure of the outer
radiative layer. 

Solving equations (\ref{eq:pressure}) and (\ref{eq:polytrope}) 
with the boundary condition $\rho=\rho_{conv}\equiv 
P_{conv}\mu/kT_{conv}$ at $r\approx R_a$ 
and using $K=P_{conv}\rho_{conv}^{-\gamma}$ one finds
\begin{eqnarray}
&& \rho(r)=\rho_{conv}\left[1+\nabla_{ad}
\frac{GM_c}{K\rho_{conv}^{\gamma-1}}\left(\frac{1}{r}-
\frac{1}{R_a}\right)\right]^{1/(\gamma-1)}\nonumber\\
&& =\rho_{conv}\left[1+\nabla_{ad}\left(1-\frac{\nabla_{ad}}{\nabla_0}
\right)^{1/(4-\beta)}
\left(\frac{R_B}{r}-
\frac{R_B}{R_a}\right)\right]^{1/(\gamma-1)}
\label{eq:conv_sol}
\end{eqnarray}
For $r\lesssim R_a(1-R_a/R_B)$ gas density strongly exceeds
$\rho_{conv}$ and one obtains
\begin{eqnarray}
\left(\frac{\rho}{\rho_{conv}}\right)^{\gamma-1}=
\frac{T}{T_{conv}}\approx\nabla_{ad}\frac{T_0}{T_{conv}}
\left(\frac{R_B}{r}-\frac{R_B}{R_a}\right)
\label{eq:conv_zero}
\end{eqnarray}
Note that this temperature profile would be identical to 
(\ref{eq:rad_zero_t}) found for envelopes with radiative 
interiors if $\nabla_{ad}$ were replaced with $\nabla_0$; 
apparently, temperature behavior is not very sensitive to
the details of physics determining the atmospheric structure.


\section{Atmospheres with outer convective zone.}
\label{sect:solutions_inner}
 

Whenever planetesimal accretion luminosity is high, 
luminosity radius $R_L$ can exceed $R_B$ in the optically thin case 
(c.f. \S \ref{subsect:opt_thin}) or $(R_B\lambda)^{1/2}$ in the 
optically thick (c.f. \S \ref{subsect:pure_dif}). Based on the 
results of \S \ref{sect:solutions} we can anticipate that 
intense energy release at the core surface would strongly 
affect gas temperature even beyond the Bondi radius,  
and  this qualitatively changes the 
atmospheric structure.

We will concentrate on the optically thick 
 case with 
\begin{eqnarray}
R_B\gg \lambda, ~~~~~R_L^2\gtrsim R_B\lambda
\label{eq:inequality3}
\end{eqnarray} 
(directly opposite to [\ref{eq:inequality2}]), 
typical in the terrestrial region (see Figure 
\ref{fig:regions}).  Since atmosphere is optically thick 
equation (\ref{eq:purely_rad_sol}) should 
determine the envelope structure if it were 
convectively stable. However, it is easy to show that 
under the conditions (\ref{eq:inequality3}) gas around the 
core cannot be convectively stable even at $R_{out}\gg R_B$!
Indeed, temperature gradient is given by (\ref{eq:nabla_in}) and 
one finds that far from the core $\nabla>\nabla_{ad}$ 
provided that
\begin{eqnarray}
\nabla_{ad}<\nabla(\infty)=
\frac{3}{4}\frac{R_L^2}{R_B\lambda}=\frac{3}{64\pi}
\frac{L\kappa_0\rho_0 c_0^2}{G M_c\sigma T_0^4}.
\label{eq:nabla_conv}
\end{eqnarray}
If (\ref{eq:inequality3}) holds, 
$\nabla\approx R_L^2/(R_B\lambda)\gg 1$ already at 
$r\sim R_{out}$, meaning that even the outer part of the 
envelope is convective. This is completely different  
from the situation typical for the region of giant planets
(see \S \ref{sect:solutions}) where 
conditions are such that protoplanetary envelopes always have 
an almost isothermal outer radiative zone. The reason for 
this difference is 
that in the present case large energy flux 
escaping from the core severely affects gas temperature 
outside Bondi sphere where gas pressure is still almost 
equal to its nebular value $P_0$. Consequently,  
temperature gradient $\partial\ln T/\partial\ln P$ takes 
on a high value outside of $R_B$ (unlike the case studied in
\S \ref{sect:solutions}) giving rise to convection.
As a result, envelope acquires an {\it outer convective zone}.

\begin{figure}
\plotone{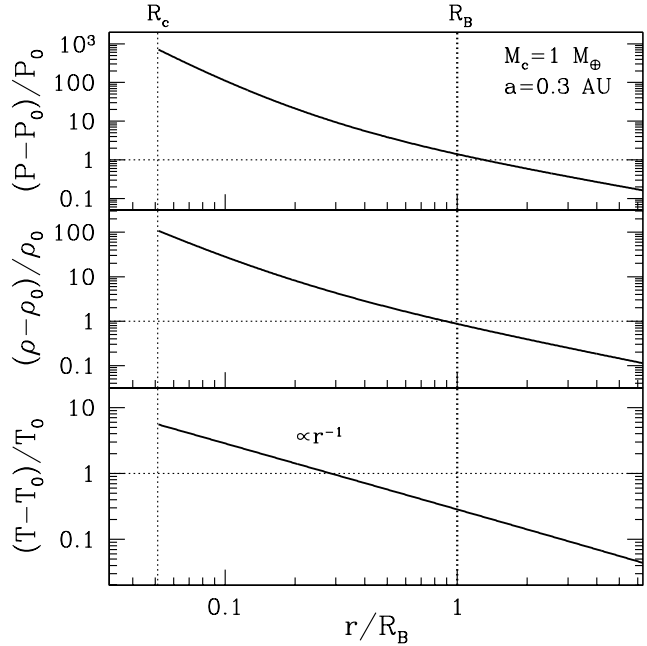}
\caption{Same as Figure \ref{fig:rad_str_5_5} but for 
$M_c=1~M_\oplus$ at $0.3$ AU. In
this particular case $P_0=3\times 10^{-3}$ bar, 
$\rho_0=1.6\times 10^{-7}$ g cm$^{-3}$ and $T_0=550$ K.  
Also $\lambda \ll R_B$ and 
$R_B\lambda/R_L^2\approx 0.06$, i.e envelope is optically 
thick everywhere and possesses an outer convective zone,
see \S \ref{sect:solutions_inner}. The interior of this
atmosphere is also convective for a particular choice 
of parameters (the same as for Figure \ref{fig:rad_str_5_5}) 
used in producing this Figure.
\label{fig:rad_str_1_0.3}}
\end{figure}

Calculation similar to that leading 
to (\ref{eq:conv_sol}) yields the following density structure 
in the convective part of atmosphere:
\begin{eqnarray}
\rho(r)=\rho_0\left[1+\nabla_{ad}
\left(\frac{R_B}{r}-
\frac{R_B}{R_{out}}\right)\right]^{1/(\gamma-1)}.
\label{eq:conv_sol_out}
\end{eqnarray}
This expression shows that gas density, pressure, and temperature 
deviate weakly from their nebular values $\rho_0$, $P_0$, and
$T_0$ as long as $R_B\lesssim r\lesssim R_{out}$, but they increase as  
power laws of $R_B/r$ inside the Bondi sphere. 
Apparently, internal atmospheric structure (for $r\lesssim R_B$) 
is rather insensitive to a particular choice of radius
$R_{out}$ at which boundary conditions (\ref{eq:bound}) are
set, as long as $R_{out}\gtrsim R_B$.

Structure of the interior regions of the envelope depends
on the opacity behavior as described in \S 
\ref{subsect:convection}: atmosphere is convective from $R_{out}$ 
all the way to $R_c$ whenever $\nabla_0>\nabla_{ad}$, but it becomes
convectively stable and radiative at some depth if $\nabla_0<\nabla_{ad}$. 
In the latter case, for the inner radiative region to 
exist it is necessary that convection stops above the core surface. 
Suppose that $\nabla_0<\nabla_{ad}$ and that transition between 
the outer convective zone and the inner radiative region takes place
at a distance $R_{rad}$ from the core center. 
Pressure, temperature, and density at this point are 
$P_{rad},~T_{rad},~\rho_{rad}$; they can be uniquely related to 
$R_{rad}$ by (\ref{eq:conv_sol_out}) since the outer boundary
of the radiative zone is also an inner boundary of the convective 
zone. Within the radiative zone behavior of $\nabla(T)$ can be
described by (\ref{eq:nabla_in_thin}) with $P_{ph}$ and $T_{ph}$ 
replaced by $P_{rad}$ and $T_{rad}$. Then one can easily fix the 
value of $R_{rad}$ from the condition $\nabla(T_{rad})=\nabla_{ad}$: 
\begin{eqnarray}
R_{rad}\approx R_B \frac{\nabla_{ad}}{\Theta},~~
\Theta\equiv\left(\frac{3}{4\nabla_{ad}}
\frac{R_L^2}{R_B\lambda}\right)^{\frac{\nabla_{ad}}
{(4-\beta)(\nabla_{ad}-\nabla_0)}}.
\label{eq:R_rad}
\end{eqnarray}
It is clear that $\Theta\gg 1$ and $R_{rad}\ll R_B$ 
because of (\ref{eq:inequality3}). Inner radiative 
zone exists only if $R_{rad}>R_c$ in addition to
$\nabla_0<\nabla_{ad}$. Temperature and density 
at $R_{rad}$ are, of course, much larger than $T_0$ and $P_0$:
$T_{rad}/T_0=(\rho_{rad}/\rho_0)^{\gamma-1}\approx \Theta$,
see (\ref{eq:conv_sol_out}). 

We also briefly discuss the atmospheric structure in the 
optically thin case\footnote{This particular relationship 
between $R_B,~R_L$, and $\lambda$ can be realized only in the
region of giant planets for small cores accreting in the fast regime, 
see Figure \ref{fig:regions}. We mention it here mainly for completeness.} 
\begin{eqnarray}
R_B\ll\lambda,~R_B\ll R_L
\label{eq:inequality4}
\end{eqnarray} 
opposite to that considered in 
\S \ref{subsect:opt_thin}. Using equation (\ref{eq:t_approx}) to 
determine the temperature structure in the optically 
thin part of the envelope one finds $T$ to be strongly perturbed 
beyond the Bondi radius because of the second inequality
in (\ref{eq:inequality4}). Then it is rather clear that
such atmospheres should possess outer convective zones,  
similar to the case when (\ref{eq:inequality3}) holds. 
This conclusion is independent 
upon the relationship between $R_L$ and $\lambda$. We do not
consider the interior structure of such envelopes in this study.


\section{Envelope mass and critical core mass.}
\label{subsect:mass}


We define mass of the envelope $M_{env}$ as 
\begin{eqnarray}
M_{env}\equiv 4\pi\int\limits_{R_c}^{r_B}\rho(r^\prime)
r^{\prime 2}dr^\prime,
\label{eq:m_env}
\end{eqnarray}
where we have chosen $R_B$ to be the outer boundary of the 
envelope. We compute $M_{env}$ 
separately for atmospheres with the outer radiative and convective 
zones since it will turn out that masses 
are very different in the two cases. Using these results 
we also estimate the critical core mass necessary for the 
initiation of a runaway gas accretion in \S \ref{subsect:crit_mass}.


\subsection{Envelopes having outer radiative zone.}
\label{subsect:mass_rad}

Results of \S \ref{sect:solutions} demonstrate that gas density 
in the atmospheres having outer radiative zone  
increases exponentially between $R_B$ and $R_a$. As a result,
most of the atmospheric mass is contained within $\sim R_a$.
In Appendix \ref{ap:mass} we demonstrate that $M_{env}$ is 
given by
\begin{eqnarray}
M_{env}\approx 4\pi\Psi_1\rho_0 R_B^3
\left(\frac{R_B\lambda}{R_L^2}\right)^{1/(1+\alpha)},
\label{eq:mass_env}
\end{eqnarray}
where $\Psi_1$ is
a weak (logarithmic) function of $R_B\lambda/R_L^2$ given 
by (\ref{eq:theta_def}).
Envelope mass is dominated by the contribution coming from 
$r\sim R_a$ in all dynamically stable atmospheres with 
convective interiors and in all atmospheres with radiative 
interiors having $\nabla_0>1/4$; in these cases the 
inner part of the 
atmosphere near the core contributes to $M_{env}$ only 
weakly, see Appendix \ref{ap:mass}. We restrict ourselves 
to studying only these two important classes of envelopes.

Of special interest is the case when atmospheric opacity is 
independent of pressure (and, consequently, density), i.e. 
$\alpha=0$. Then, regardless of whether envelope interior is 
radiative or convective, one finds using  (\ref{eq:Bondi}), 
(\ref{eq:lambda}), (\ref{eq:R_L}), 
(\ref{eq:R_a}), and (\ref{eq:m_dot}) that
\begin{eqnarray}
&& M_{env}=64\pi^2 \Psi_1\left(\frac{GM_c\mu}{k}\right)^4
\frac{\sigma}{\kappa_0 L}\nonumber\\
&& \approx
\left(\frac{M_c}{M_\oplus}\right)^{8/3}
\frac{0.1~\mbox{cm$^2$g$^{-1}$}}{\kappa_0}
\left\{
\begin{array}{l}
8\times 10^{27}~\mbox{g}~a_{10}^{3},~~~~\mbox{slow},\\
6\times 10^{24}~\mbox{g}~a_{10}^{2},~~~~\mbox{int.}, \\
2\times 10^{23}~\mbox{g}~a_{10}^{3/2},~~~\mbox{fast}.
\end{array}
\right.
\label{eq:indep}
\end{eqnarray}
Different numerical estimates are for three accretion regimes 
described in Appendix \ref{ap:acc_rate}
and assume molecular gas of cosmic composition, $\beta=1$, 
and $\gamma=7/5$ (convective interior).
The peculiarity of this special case is that envelope mass
is virtually independent of the temperature and density 
in the surrounding nebula. Indeed,
both $T_0$ and $\rho_0$ enter (\ref{eq:indep}) only 
logarithmically through the dependence of $\Psi_1$ on 
$R_B/R_a$, see (\ref{eq:R_a}) and (\ref{eq:theta_def}). 
Local conditions in the protoplanetary disk affect 
$M_{env}$  only through gas 
opacity $\kappa_0$ and planetesimal accretion luminosity 
$L$. This was first noticed by Stevenson (1982) who 
discovered this feature while studying radiative envelopes with 
constant gas opacity, $\alpha=\beta=0$.
But equation (\ref{eq:indep}) demonstrates that insensitivity of 
$M_{env}$ to $T_0$ and $\rho_0$ is a more general
phenomenon since it holds even 
when atmospheric opacity varies with temperature and for 
convective as well as radiative interiors, provided only that
$\alpha=0$, i.e. opacity is independent of the gas density.

When $\kappa$ does depend on 
pressure ($\alpha\neq 0$), one finds that
\begin{eqnarray}
M_{env}\propto\left[\frac{\rho_0^\alpha(M_c\mu)^{4+3\alpha}}
{T_0^{3\alpha}\kappa_0 L}\right]^{1/(1+\alpha)},
\label{eq:dep}
\end{eqnarray}
i.e. envelope mass depends on nebular properties in
this more general case. But even then the 
detailed character of dependence is not determined 
by whether atmospheric interior is radiative or 
convective, but only by the opacity dependence on $P$. 

It might seem surprising that mass of the envelope 
with convective interior can be sensitive to the gas opacity 
since the energy transfer below $R_a$ is not done by 
radiation. The explanation lies
in the presence of the outer radiative zone 
above $R_a$ which is non-adiabatic. Because of that 
gas entropy at the edge of the convective zone
(at $\approx R_a$) is set by the radiative energy transfer in the
outer atmosphere and depends on the gas opacity $\kappa_0$
(see equation [\ref{eq:K}]).
This is the origin of dependence of $M_{env}$ on
the nebular opacity in the case of atmospheres with convective
interiors.

\begin{figure}
\plotone{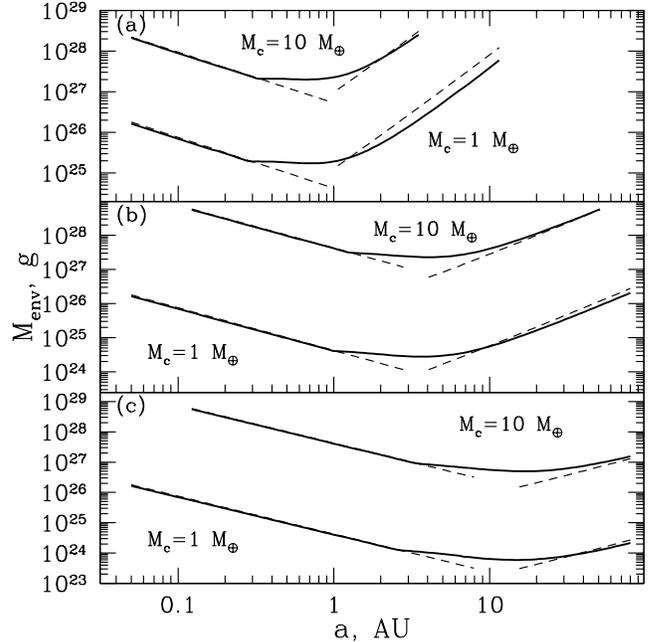}
\caption{Mass of the envelope for two values of 
$M_{core}=1,~10~M_\oplus$ as a function of semi-major 
axis $a$ in the nebula. Different panels correspond to different 
planetesimal accretion regimes: (a) slow, (b) intermediate, and 
(c) fast. Solid curves labeled with corresponding core masses
represent results of numerical calculation of $M_{env}$. 
Dashed lines display analytical estimates of $M_{env}$ 
for the same $M_c$ given by (\ref{eq:indep}) and 
(\ref{eq:mass_conv}) for large and small $a$ respectively.
\label{fig:env_mass}}
\end{figure}

In Figure \ref{fig:env_mass} we plot $M_{env}$ for  
cores of specified mass ($1$ and $10$ M$_{\oplus}$) as a function
of core's distance from the Sun $a$ for 
fast, intermediate, and slow planetesimal 
accretion regimes (see Appendix \ref{ap:acc_rate}). 
Numerical results of the envelope structure calculations for
$\alpha=0, \beta=1$, $\kappa_0=0.1$ cm$^2$ g$^{-1}$, $\gamma=7/5$ 
are shown by solid curves; 
dashed line at large $a$ is our analytical estimate of $M_{env}$
given by (\ref{eq:indep}). As Figure \ref{fig:env_mass}a 
demonstrates, formula (\ref{eq:indep}) somewhat overestimates 
$M_{env}$: because of the finite size of the core our extension 
of integration in (\ref{eq:theta_def}) to 
zero (instead of $R_c$) leads to the  overestimate
of $M_{env}$; relative correction is $\sim(R_c/R_a)^{1/2}$
which is small but sometimes non-negligible. We checked that 
this discrepancy goes away when we artificially set $R_c$ to 0 
in our numerical calculations.


\subsection{Envelopes having outer convective zone.}
\label{subsect:mass_conv}

Mass of the atmosphere possessing outer convective zone
is calculated using 
density profile (\ref{eq:conv_sol_out}). Assuming  
that (\ref{eq:conv_sol_out}) holds up to the core's surface 
(fully convective envelope, $\nabla_0>\nabla_{ad}$) one 
finds
\begin{eqnarray}
&& M_{env}=4\pi \Psi_2\rho_0 R_B^3\approx 4\times 10^{24}
~\mbox{g}~\left(\frac{M_c}{M_\oplus}\right)^3 a_1^{-5/4},
\nonumber\\     && 
\Psi_2(\gamma)
\equiv \int\limits_0^1 z^2\left(1+
\frac{\nabla_{ad}(\gamma)}{z}\right)^{1/(\gamma-1)}dz,
\label{eq:mass_conv}
\end{eqnarray}
where we have assumed $R_B/R_{out}\lesssim 1$. 
Numerical estimate is done for $\gamma=7/5$ ($\Psi_2\approx 0.88$).
Similar estimate of $M_{env}$ can be found in Wuchterl (1993). 

A remarkable property of atmospheres having outer 
convective zone is that gaseous mass 
contained within a Bondi sphere around the core is of 
the order of the mass of nebular gas with $\rho=\rho_0$
contained within the same volume! This is very different from 
atmospheres possessing outer radiative region, which
not only have envelope mass much higher than 
$(4/3)\pi\rho_0 R_B^3$, but also contain this mass 
within smaller volume than that of the Bondi sphere, see 
(\ref{eq:mass_env}). In addition, $M_{env}$ given by 
(\ref{eq:mass_conv}) is completely independent of the core 
luminosity and gas opacity, very much unlike the case 
studied in \S \ref{subsect:mass_rad}. 

Most of the mass in fully convective atmospheres is 
concentrated in their outer part (because of the requirement
$\gamma>4/3$ necessary for dynamical stability), which 
allowed us to set the lower integration limit in the definition 
of $\Psi_2$ to zero.
One can demonstrate that in the case of envelopes having 
inner radiative region (those with $\nabla_0>\nabla_{ad}$)
formula (\ref{eq:mass_conv}) continues to correctly 
describe the mass of the
envelope, provided that $\nabla_0\ge 1/4$ in its
radiative part (because most of the mass in radiative region 
is then concentrated near its outer edge); also, inner radiative 
region typically has rather small radial extent, see (\ref{eq:R_a}). 

In Figure \ref{fig:env_mass} analytical estimate 
(\ref{eq:mass_conv}) is displayed by the dashed line in the inner 
part of the nebula (terrestrial planet region, $\lesssim 1$ AU).
One can see very good agreement of analytical result with 
the predictions of more detailed numerical calculations. 

\begin{figure}
\plotone{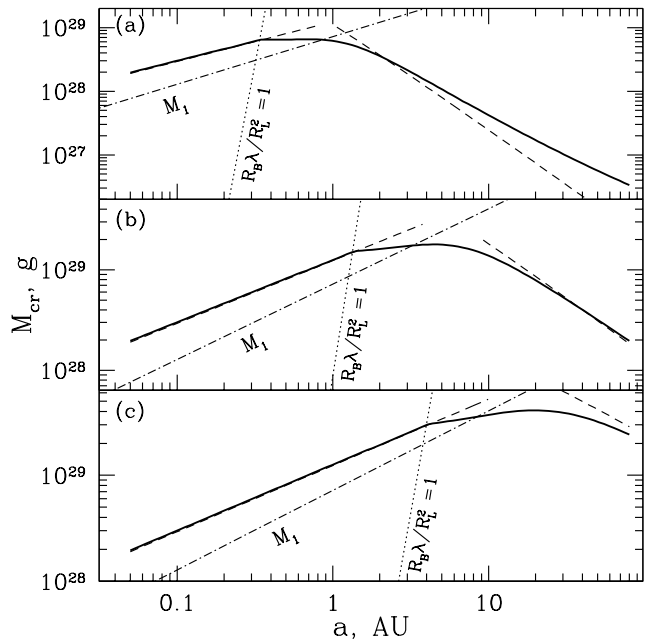}
\caption{Critical core mass as a function of semi-major axis $a$ 
in the nebula for (a) slow, (b) intermediate, and (c) fast 
planetesimal accretion rates. Solid curve represents numerical 
results, while dashed lines display estimates of $M_{cr}$ given by
(\ref{eq:m_cr_rad}) and (\ref{eq:m_cr_conv}) at large and 
small $a$ correspondingly. Dot-dashed line shows the run of fiducial 
mass $M_1$ (equation [\ref{eq:M_1}]) with $a$.
\label{fig:crit_mass}}
\end{figure}


\subsection{Critical core mass.}
\label{subsect:crit_mass}

As we have mentioned in the Introduction, both numerical calculations 
and analytical arguments suggest that phase of rapid gas accretion
onto the protoplanetary core initiates when $M_{env}\sim M_c$. 
The exact ratio of the two masses at the onset of instability 
is uncertain and can be determined only after the envelope's self-gravity 
is self-consistently taken into account which is beyond the scope 
of this study. Here we simply assume this critical ratio to be a 
free parameter $\eta\sim 1$, so that
\begin{eqnarray}
M_{env}(M_{cr})=\eta M_{cr},
\label{eq:eta}
\end{eqnarray}
where $M_{cr}$ is the critical core mass at the onset of rapid
gas accretion. For a fixed $\eta$ instability condition (\ref{eq:eta}) 
can be viewed as an equation for $M_{cr}$.

Using (\ref{eq:luminosity}), (\ref{eq:indep}), and (\ref{eq:m_dot})
we find that for atmospheres having outer radiative region 
(\S \ref{sect:solutions}) critical core mass is given by
\begin{eqnarray}
&& M_{cr}=\left[\frac{\eta\theta}{64\pi^2\Psi_1}\frac{\Omega \Sigma_p a\kappa}
{\sigma G^3M_{\odot}^{1/3}}\left(\frac{k_0}{\mu}\right)^{4}\right]^{3/5}
\nonumber\\     && \approx  \eta_{0.3}^{3/5}
\left(\frac{\kappa_0}{0.1~\mbox{cm$^2$g$^{-1}$}}\right)^{3/5}
\left\{
\begin{array}{l}
2.4\times 10^{27}~\mbox{g}~a_{10}^{-9/5},~~~~\mbox{slow},\\
1.9\times 10^{29}~\mbox{g}~a_{10}^{-6/5},~~~~\mbox{int.}, \\
1.5\times 10^{30}~\mbox{g}~a_{10}^{-9/10},~~~\mbox{fast},
\end{array}
\right.
\label{eq:m_cr_rad}
\end{eqnarray}
where $\eta_{0.3}\equiv \eta/0.3$ and $\theta$ is a parameter 
determining the efficiency of accretion. Apparently,  
core instability sets in at lower $M_c$ in the more distant 
parts of the nebula because of the rapid 
decrease of planetesimal accretion rate $\dot M$ with $a$. 
Note a strong dependence of $M_{cr}$ on the mean molecular weight 
$\mu$ (Stevenson 1982). Convective erosion of the core and 
dissolution of infalling planetesimals in the 
envelope might increase $\mu$ and considerably lower $M_{cr}$, 
facilitating rapid gas accretion. 

Opacity determining $M_{env}$ and $M_{crit}$ in (\ref{eq:mass_env}) 
and (\ref{eq:m_cr_rad}) is the opacity {\it in the outer radiative zone 
only} --- it is the radiation transfer in this region that determines 
the conditions in the innermost part of the envelope and, hence, its mass. 
Temperature in the radiative zone does not deviate strongly 
from $T_0$ and we can be more confident that the opacity behavior in this 
region can be represented by a simple
power law dependence (\ref{eq:opacity}). This assertion becomes
even more robust if opacity there is also independent of gas density 
(since density strongly varies within the radiative zone), and this is 
the case when opacity is dominated by dust absorption. 

In the case of atmosphere having outer convective zone 
(\S \ref{sect:solutions_inner}) we find after substituting 
(\ref{eq:mass_conv}) into (\ref{eq:eta}) that
\begin{eqnarray}
M_{cr}=c_0^3\left(\frac{\eta}{4\pi \Psi_2 \rho_0 G^3}\right)^{1/2}
\approx 1.2\times 10^{29}~\mbox{g}~\eta_{0.3}^{1/2}a_1^{5/8}. 
\label{eq:m_cr_conv}
\end{eqnarray}
This critical mass decreases as one goes {\it inward} in the nebula 
(unlike [\ref{eq:m_cr_rad}])
because gas density in the disk rapidly increases inward. Introducing
the Toomre stability parameter $Q\equiv \Omega c_0/(\pi G\Sigma_g)$
one can rewrite (\ref{eq:m_cr_conv}) as
$M_{cr}=(Q/4\Psi_2)^{1/2}M_1$, see definition (\ref{eq:M_1}). 
This implies that $M_{cr}\gtrsim M_1$ in gravitationally stable 
protoplanetary disks\footnote{Similar argument was advanced 
in Ikoma \etal (2001).} with $Q\gtrsim 1$. It is then clear
that $M_{env}$ given by (\ref{eq:m_cr_conv}) is only a rough 
estimate of the critical core mass because nebula cannot be 
considered static and homogeneous on the scale of $R_B$ 
(as we always assumed) when $M_c>M_1$, see 
\S \ref{sect:length_scales}.

Comparing (\ref{eq:mass_env}) with (\ref{eq:mass_conv}) we see
that presence of the outer radiative zone, requiring 
$R_B\lambda\gg R_L^2$, increases $M_{env}$ and lowers $M_{cr}$. 
This happens because radiative diffusion across almost isothermal 
outer radiative zone significantly reduces the entropy of the 
inner envelope. Lower entropy means higher density resulting in 
a higher $M_{env}$ compared to the mass which an isentropic 
atmosphere would have for the same core mass. For 
instance, in the optically thick atmosphere with 
$R_B\lambda\gg R_L^2$ the value of adiabatic constant $K$ at 
$R_a$ is
\begin{eqnarray}
K\sim K_0\left(
\frac{R_L^2}{R_B \lambda}\right)^{(\gamma-1)/(1+\alpha)},
\label{eq:K}
\end{eqnarray}
see [\ref{eq:T_conv}], [\ref{eq:P_conv}], and [\ref{eq:conv_sol}]; 
$K_0\equiv P_0\rho_0^{-\gamma}$ is the adiabatic constant of 
gas in the surrounding nebula. It follows from (\ref{eq:K}) that 
$K\ll K_0$, i.e. gas entropy in the inner atmosphere is 
{\it much lower} than it is in the surrounding nebula.  
Consequently, $M_{cr}$ for an atmosphere with the outer radiative zone is
lower than it would have been if atmosphere were isentropic (i.e. had
outer convective zone). 

In Figure \ref{fig:crit_mass} we plot the critical core mass
as a function of $a$ for different planetesimal 
accretion regimes, self-consistently 
taking into account transition between different types 
of atmospheres in the inner and outer parts of protoplanetary disk.
In addition to the results of numerical calculations of the 
atmospheric structure  
we also plot analytical approximations for $M_{cr}$
given by (\ref{eq:m_cr_conv}) in the region of terrestrial 
planets and by (\ref{eq:m_cr_rad}) in the region of giant planets.
In the terrestrial planet region theory and numerical calculations 
agree very well and one can see that $M_c> M_1$ in agreement 
with what we have said before ($Q\gtrsim 30$ at $a\lesssim 1$ AU 
for the MMSN parameters given by [\ref{eq:MMSN}]). In the giant 
planet region there are discrepancies between the theory and 
numerical results at large $a$ which are especially pronounced in Figure 
\ref{fig:crit_mass}a. They appear because of the finite size of 
the core which causes  (\ref{eq:indep}) to overestimate $M_{env}$, 
see \S \ref{subsect:mass_rad}; in the more careful numerical 
calculation a bigger core is required to trigger the runaway 
gas accretion.


\section{Discussion.}
\label{sect:disc}


One of the major results of our study is a clear 
distinction between protoplanetary atmospheres
having convective outer region which smoothly merges with
the surrounding nebula and atmospheres having radiative
zone separating dense and hot interior from the nebula. 
A specific type of atmosphere around a particular 
protoplanetary core is determined by the relationships between 
three important length scales --- $R_B$, $R_L$, and 
$\lambda$ --- provided that core mass satisfies the 
mass constraint (\ref{eq:cond}). Whenever planetesimal 
accretion rate of the core is low, meaning that $R_L$ is small
(i.e. either [\ref{eq:inequality2}] or  [\ref{eq:inequality1}] is 
fulfilled), gas temperature can be appreciably affected only deep 
inside the Bondi sphere while the pressure starts to vary 
already at the Bondi radius. As a result, according to the 
Schwarzschild criterion (\ref{eq:stab_cond}), outer parts of 
the envelope are convectively stable and energy 
is carried away by radiation. In the opposite case 
of very high accretion rate and large $R_L$ (i.e. either 
[\ref{eq:inequality3}] or [\ref{eq:inequality4}] holds), 
gas temperature is perturbed even outside the Bondi sphere, and 
condition (\ref{eq:stab_cond}) predicts that gas is 
convectively unstable for $r\gtrsim R_B$, where pressure
perturbation is small. Moreover, in the latter case 
even if the innermost parts of the envelope tend 
to settle onto the radiative (convectively stable) 
configuration, they would be able to switch to a 
radiative solution only very deep in the envelope, 
see equation (\ref{eq:R_rad}). 

This segregation of atmospheres  into two major classes
depends to some extent on whether the outer parts of the 
atmosphere are optically thick or thin. 
For example, envelope structure in the optically thick 
case depends only on the value of the dimensionless 
parameter $\lambda R_B/R_L^2$ --- roughly the inverse of 
the radiative temperature gradient far from the core
(see [\ref{eq:Bondi}], [\ref{eq:lambda}], [\ref{eq:R_L}], 
and [\ref{eq:defs}]), and is completely insensitive to 
the individual relationships between $\lambda$, $R_L$, and 
$R_B$ (as long as $\lambda\ll R_B$): envelope has outer convective 
zone when $\lambda R_B/R_L^2\lesssim 1$ and outer radiative 
zone when $\lambda R_B/R_L^2\gtrsim 1$. Analogous condition 
was formulated by Wuchterl (1993) in terms of nebular density
$\rho_0$. Separation between the two classes of atmospheres
in the optically thin case is somewhat more complicated (e.g. see 
constraint [\ref{eq:inequality1}]) and one has to 
pay attention to the individual relationships between 
the three length scales. Results of \S \ref{sect:solutions} \& 
\ref{sect:solutions_inner} cover all such 
possibilities. 

Because these characteristic length scales vary with the distance from
the Sun, a specific type of atmosphere forming around the core of
a given mass depends on $a$. This is caused primarily by the variation of the
planetesimal accretion rate throughout the protoplanetary disk: 
in the terrestrial region the planetesimal surface density is high 
while the dynamical timescale is short leading to high $\dot M$,
large $R_L$, and small $R_B\lambda/R_L^2$. Another important factor
is a steep dependence of $\lambda$ on $a$ 
(see Figure \ref{fig:scales}). As a result, atmospheres around 
massive cores in the inner parts of protoplanetary disks 
(within roughly $0.5-2$ AU depending on the planetesimal accretion 
regime) where $R_B\lambda\lesssim R_L^2$ possess outer 
convective zones, see Figure \ref{fig:regions}a. 
On the contrary, in the region of giant planets both $\Sigma_p$ 
and $\Omega$ are small, which
decreases $\dot M$ making $R_L$ small as well. Photon mean free path
there is large because gas density is very low (disk
can be optically thin). These factors conspire to rapidly increase
$R_B\lambda/R_L^2$  with $a$ and allow cores outside of 
$\sim (0.5-2)$ AU to have massive atmospheres with quite 
extended outer radiative zones, see Figures \ref{fig:rad_str_5_5} 
\& \ref{fig:rad_str_10_30}. Some exceptions are possible (e.g. 
cores having $R_B\ll \lambda$ and $R_B\ll R_L$ in the region 
of giant planets have convective outer atmospheres, 
see \S \ref{sect:solutions_inner} and Figure \ref{fig:regions}) 
but they typically occur for 
rather small cores unable of retaining massive atmospheres.


\subsection{Comparison with previous studies.}
\label{subsect:comparison}

Envelopes with the outer convective region have been 
previously considered by Perri \& Cameron (1974),
Wuchterl (1993), and Ikoma \etal (2001), who found the 
masses of such envelopes to be 
rather small which translated into large critical core 
mass. In addition, properties of 
such fully convective atmospheres were found to strongly 
depend on the temperature and density in the surrounding 
nebula. This is exactly the picture that we described in \S 
\ref{sect:solutions_inner}. Wuchterl (1993) suggested that the
sensitivity to external conditions is a consequence of 
envelopes being fully convective, but as we demonstrate in \S 
\ref{sect:solutions_inner} what is really important is 
the convection {\it in the outermost region of atmosphere} 
only, independent of whether the atmospheric interior 
is convective or radiative. Presence of the outer convective 
region sets entropy of the inner atmosphere equal to 
the entropy of the nebular gas, so that variations in
the external conditions directly affect the overall structure 
(and mass) of the envelope. Nebular entropy is quite high, 
which makes atmospheres not very dense and accounts 
for the low masses of such envelopes, see Figure 
\ref{fig:env_mass}. As our analysis demonstrates, 
Perri \& Cameron (1974) and Wuchterl (1993) have probably 
stretched their assumptions too far by calculating 
$M_{cr}\approx 60$ M$_\oplus$ for convective
envelopes at $5$ AU, in the region of giant planets, where 
atmospheres should in fact have outer radiative (not 
convective) zones.

Envelopes possessing a radiative region between the inner, 
dense parts of the atmosphere and the nebular gas outside have first  
been studied numerically by Harris (1978), Mizuno \etal 
(1978), and Hayashi \etal (1979). Mizuno (1980) and Stevenson (1982) 
were the first to notice that presence of the outer radiative region 
makes critical core mass virtually independent upon the density 
and temperature of the surrounding gas, provided that the opacity in the 
outer atmosphere is {\it constant}, i.e. $\kappa$ is independent of 
either gas pressure or temperature ($\alpha=\beta=0$). 
This is completely different from envelopes having outer 
convective region and occurs because outer radiative zone 
decouples inner atmosphere from the surrounding nebula. As we 
demonstrated in \S \ref{subsect:mass_rad} this insensitivity of 
$M_{cr}$ to the external nebular conditions  
holds also in a more general case of $\alpha=0$ and arbitrary 
$\beta$ (opacity independent on the gas density), see equation 
(\ref{eq:indep}). This scaling is typical for the dust opacity which
should dominate\footnote{Unless the dust opacity is very small, 
$\lesssim 10^{-2}$ cm$^2$ g$^{-1}$} over the molecular opacity 
due to H$_2$ and H$_2$O under the conditions typical in the 
outer radiative zone, see opacity plots in Hayashi \etal (1979) 
and Mizuno (1980) and Figures \ref{fig:rad_str_5_5} \& 
\ref{fig:rad_str_10_30}. Thus, one would naturally expect envelope 
mass and critical core mass to be independent of $\rho_0$ and $T_0$ 
(except for the local value of opacity $\kappa_0$ which may scale 
with local temperature $T_0$); this result is also completely 
insensitive to the structure of the innermost part ($r\lesssim R_a$) 
of the envelope, be it radiative or convective. Of course, as 
soon as $\alpha$ is nonzero, this degeneracy breaks and critical 
core mass starts to depend on $T_0$ and $\rho_0$, as equation 
(\ref{eq:dep}) demonstrates.

Previous studies self-consistently accounting 
for the presence of the outer radiative zone 
(Hayashi \etal 1979; Mizuno 1980; Stevenson 1982; 
Nakazawa \etal 1985; Ikoma \etal 2000) have 
found $M_{cr}\sim 10$ M$_\oplus$, depending on 
the dust opacity and accretion luminosity of the core. These 
authors typically assumed constant 
$\dot M$, i.e. $M_c\propto \tau$, while we consider several 
possible accretion regimes 
taking into account the dependence of $\dot M$ on the core mass and its 
distance from the Sun. Note that for the same 
core formation timescale our assumed accretion law $M_c\propto \tau^3$ 
(see Appendix \ref{ap:acc_rate}) yields higher $\dot M$ in the end 
of core accretion than constant $\dot M$ does; 
this acts to increase $M_{cr}$.
If nucleated instability sets in when 
envelope mass is $30\%$ of the core mass and 
$\kappa_0=0.1$ cm$^2$ g$^{-1}$, as we assumed 
in all our numerical estimates, we find that critical core mass in the 
fast regime of planetesimal accretion is $53$ M$_\oplus$ at 
$5$ AU (present Jupiter's location) and $62$ M$_\oplus$ at 
$10$ AU (current Saturn's location). In the intermediate 
accretion regime $M_{cr}$ is $30$ M$_\oplus$ and $23$ M$_\oplus$ at $5$
and $10$ AU correspondingly; 
it drops to $1.8$ M$_\oplus$ and $0.7$ M$_\oplus$ at these 
locations in the slow accretion regime. Ikoma \etal (2000) have found 
values of $M_{cr}$ that are somewhat lower (sometimes by $\sim 2$) 
than those found in this work for the same $\dot M$ and $\kappa_0$ because
they used opacity in the outer atmosphere independent of $P$ and 
$T$ which has effect of increasing $M_{env}$ (see below) while we 
use realistic dust opacity. Our neglect of atmospheric self-gravity is 
another reason for this difference.

Inner envelope ($r\lesssim R_a$) can be either radiative or 
convective, depending on the 
detailed opacity behavior. We found that this part of atmosphere 
is radiative provided that $\alpha$ and $\beta$ in (\ref{eq:opacity})  
are such that the condition (\ref{eq:cond_rad}) is fulfilled; 
if $\nabla_0>\nabla_{ad}$ 
the inner envelope must be convective. Stevenson (1982) used 
constant opacity in his study which made the entire envelope 
convectively stable and energy was transferred solely by radiation. 
On the other hand, in the important case of opacity dominated by 
small dust grains one would take 
$\alpha=0$ and $\beta\approx 1-2$. Equation (\ref{eq:cond_rad}) 
demonstrates that envelopes with such opacity should be 
convectively unstable if they consist of diatomic gas with 
$\gamma=7/5$, see Figures \ref{fig:rad_str_5_5} \& 
\ref{fig:rad_str_10_30}. 

Of course, dust opacity cannot dominate in the
whole envelope --- at large depth pressure and temperature are  
so high that H$_2$ and H$_2$O opacities become more important, and
their behavior cannot be described by a simple power law dependence 
(\ref{eq:opacity}). Hayashi \etal (1979) and Mizuno (1980) showed that
envelope is typically {\it convectively unstable} when these molecular opacities 
dominate. In addition to molecular $\kappa$ these authors also included 
constant dust opacity similar to Stevenson (1982), which had the effect of
making atmosphere radiative in the outer region where $\kappa$ was 
dominated by dust and convective at greater depth where molecular 
opacity was more important. In their case this transition typically 
occurred quite deep in 
the envelope, way below what we would call $R_a$. Presence of such 
extensive radiative zone increases $M_{env}$ and decreases $M_{cr}$: as long as 
$\alpha=\beta=0$, density in the radiative region varies as $r^{-3}$ 
and atmospheric mass is evenly distributed in equal logarithmic intervals 
in $r$, see Appendix \ref{ap:mass}. This augments total envelope mass 
compared to the mass contained in just the outermost part near $R_a$
by additional factor equal to the logarithm of ratio of the outer 
to inner radii of this radiative zone (and this ratio is large). 
But if one uses opacity with $\nabla_0>\nabla_{ad}$ then there 
is no radiative region below $R_a$, envelope becomes convective at much smaller 
depth, right below $R_a$, and most of the atmospheric mass 
is concentrated near $R_a$, see Appendix \ref{ap:mass}. We expect 
the latter to occur [and $M_{cr}$ to be higher than what
Hayashi \etal (1979), Mizuno (1980) and Ikoma \etal (2000) 
have found] whenever opacity is dominated 
by small dust grains, since $\alpha=0$ and $\beta\approx 1-2$ in this case. 
Convection at small depth in the case of dust opacity 
also wipes out the need to know the molecular 
opacities deep inside the atmosphere very accurately: behavior of $\kappa$
there is irrelevant as long as envelope is convective below $R_a$.


\subsection{Critical mass: implications for planet formation.}
\label{subsect:core_mass}

Our results for the critical core mass necessary to trigger the 
nucleated instability fall within the range of previously
estimated values of $M_{cr}$. In the terrestrial planet region, 
at 1 AU, we estimate\footnote{Accurate value of $M_{cr}$ in 
the terrestrial region can only be 
obtained after the effects of the vertical disk structure, 
differential shear, complicated flow geometry within the Hill 
sphere, etc. are properly included because 
in this part of protoplanetary disk $M_c>M_1$ (see 
\S \ref{sect:length_scales} and Figure 
\ref{fig:crit_mass}).} 
$M_{cr}\approx 20$ M$_\oplus$ if planetesimal 
accretion is in the intermediate or fast regime:
$R_B\lambda/R_L^2\lesssim 1$ in both cases, 
meaning atmosphere with outer convective zone. 
Perri \& Cameron (1974) and Wuchterl (1993), who studied 
convective envelopes, would have obtained the same value of
$M_{cr}$ if they were to calculate it at 1 AU. 

At the same time, accretion in the intermediate or fast regime
which keeps $R_L$ high can only proceed for rather limited time 
span at $\sim 1$ AU until the isolation mass is reached (Lissauer 1993). 
Beyond this point protoplanetary cores accrete basically at 
their geometric cross-section, i.e. in the slow accretion regime 
(Appendix \ref{ap:acc_rate}; Goldreich \etal 2004), for which 
$R_B\lambda/R_L^2$ could be somewhat higher than 1. 
After this transition occurs, critical 
core mass at $1$ AU goes down to about $10$ M$_\oplus$ 
(see Figure \ref{fig:crit_mass}).
Inward from $1$ AU, in the region where ``hot Jupiters'' were 
discovered, $M_{cr}$ decreases 
and reaches $\approx 5$ M$_{\oplus}$ at $0.1$ AU. Since 
the amount of refractory material in the inner parts of
protoplanetary disks is rather small  and it takes very long time
($\sim 10^8$ yr, see Chambers 2001) to collect this material into 
$\sim 1$ M$_\oplus$ 
protoplanets, we conclude that cores forming in situ in the 
terrestrial zone would not be able to undergo nucleated 
instability before gas in the nebula has gone away (within 
$\sim 10^7$ yr). These considerations strengthen the argument 
according to which ``hot Jupiters'' or their massive progenitor 
cores have migrated to their present locations from elsewhere.

Although masses of existing terrestrial planets are clearly 
too low to drive the nucleated instability, they were high enough 
to retain quite substantial atmospheres while the nebular gas was 
still around. Our calculations demonstrate that 
$M_{env}=4\times 10^{21}$ g, $10^{25}$ g, $3\times 10^{25}$ g,
and  $3\times 10^{22}$ g for Mercury, Venus, Earth, and Mars 
correspondingly. This is to be compared with the present 
atmospheric masses of these planets: no atmosphere on Mercury, 
$5\times 10^{23}$ g, $5.2\times 10^{21}$ g, and $6.5\times 10^{18}$ g 
on Venus, Earth, and Mars. Clearly, the primaeval atmospheres of 
terrestrial planets have been heavily depleted. Massive primordial 
atmosphere can cause severe blanketing and melting of the 
core surface, effect which has been first considered by 
Hayashi \etal (1979).

As we demonstrated in \S \ref{subsect:comparison} critical core 
mass in the region of giant planets sensitively depends on the 
accretion regime: at $10$ AU it is less than M$_\oplus$ in the 
slow regime and as large as $60$ M$_\oplus$ in the 
fast. At the same time,  present mass of the Jupiter's 
solid core is estimated to be $\lesssim 10$ M$_\oplus$; 
current Saturn's core mass is between $10$ M$_\oplus$ and 
$25$ M$_\oplus$ (Saumon \& Guillot 2004). The initial core 
masses of these two planets could have been higher ---
their total masses in high-Z elements can be as high as 
$30-40$ M$_\oplus$ because some refractory materials can be 
dissolved in their envelopes. Transporting these elements from the 
core into the envelope would require some quite efficient mixing
process such as vigorous convection 
(Stevenson 1982) and it is not clear at present if this 
is possible. 

Accumulation of the solid cores in the giant planet region 
must have proceeded in the 
intermediate or fast accretion regime (or some combination 
of the two) because smaller $\dot M$ would not
allow solid cores to form before the nebula dissipation 
this far from the Sun. This translates into values of $M_{cr}$
which are higher than the present core masses of Jupiter and 
Saturn. Thus, current data suggest that if cores were not 
initially more massive and 
subsequently eroded they were not large enough to 
trigger nucleated instability and accrete gas. The discrepancy 
is especially dramatic for Jupiter with its very low $M_c$. 
One way to resolve this paradox is to hypothesize that  
nebular opacity $\kappa_0$ was lower than $0.1$ cm$^2$ g$^{-1}$
which brings $M_{cr}$ down, see (\ref{eq:m_cr_rad}). 
Another possibility is suggested by the observation 
that core formation in the
intermediate or fast regime can take rather short time, of order
several Myrs, see (\ref{eq:t_acc}). In this case, after the
core of Jupiter has been mostly formed, there was still enough  
gas around to accumulate atmosphere. At this point, since the 
isolation mass has been reached, planetesimal accretion 
became considerably slower 
meaning that $M_{cr}$ went down dramatically, below 
already accumulated $M_c$. As a result, gaseous envelope could not 
be supported by the energy release due to the significantly lowered 
$\dot M$ and it started to slowly contract on a thermal 
timescale.  Thus, the transition between 
different planetesimal accretion regimes must be accompanied 
by the period of envelope adjustment similar to that
described in Introduction. After $M_{env}$ reached $\sim M_c$ runaway 
gas accretion commenced. Similar scenario was studied
numerically by Pollack \etal (1996) and Ikoma \etal 
(2000). We envisage the same picture to hold for Saturn
as well, although its core must have taken longer to form 
than that of Jupiter and the nebula was appreciably depleted
by that time; this may account for the smaller gaseous mass 
of Saturn. 

Assembly of the cores of ice giants --- Uranus
and Neptune --- would have required even more time. 
Relatively small atmospheres of these planets containing 
only $1-4$ M$_\oplus$ of H and He suggest that they either 
never underwent nucleated instability, or if they did this
happened only after nebula has been very strongly depleted.
In the former case all the gas that we see now in Uranus
and Neptune has come from steady state 
atmospheres around the cores which were less massive than 
present cores of ice giants; after nebula went away these 
(sub-isolation mass) cores merged retaining their gaseous 
content (Genda \& Abe 2003, 2004) and producing Uranus and Neptune. 
In the latter case fast accretion leads to the accumulation of 
isolation mass (essentially the present day mass of ice giants) 
and to subsequent nucleated instability before the nebula 
dispersal; however, if at that moment nebula was depleted by less 
than $10^2-10^3$, Neptune and 
Uranus would have much higher gaseous masses than they do now. 

Critical mass rather strongly depends on the envelope 
composition, namely on $\mu$. Stevenson (1982) was the 
first to notice this fact for purely radiative atmospheres. 
Our equation (\ref{eq:m_cr_rad}) confirms and generalizes this
observation --- we find $M_{cr}\propto \mu^{-12/5}$ for
{\it any} envelope with the outer radiative zone (independent of
whether its interior is radiative or convective). Dissolution
of infalling planetesimals, erosion of the core by vigorous 
convection, or evaporation of some volatile icy content of the 
core can increase $\mu$ and decrease $M_{cr}$ considerably
("superganymedean puffballs", Stevenson 1984). On the other hand, 
envelope enrichment in high-Z elements might also increase 
opacity in the outer atmosphere and this can at least partially 
counteract the decrease of $M_{cr}$ due to large $\mu$.


\subsection{Validity of assumptions.}
\label{subsect:assumptions}

Because of the inherently analytical nature of this study aimed 
at singling out the most important aspects of 
atmospheric structure we have neglected a number of 
relevant phenomena which may be important in some
cases. Among them are the dissolution of infalling 
planetesimals (Pollack \etal 1996), 
opacity jumps due to dust grain melting (Mizuno 1980), 
increase of planetesimal capture cross-section caused 
by the presence of the atmosphere (Inaba \& Ikoma 2003), etc. 
We have also employed a set of simplifying
assumptions such as hydrostatic and thermal equilibrium of  
atmosphere, negligible gas accretion luminosity, and so on. 
In Appendix \ref{ap:thermal_time} we 
demonstrate these assumptions to be valid. Our treatment 
of convection relies on the absence of entropy gradient 
in the convective regions and in Appendix  \ref{ap:conv_efficiency} 
we checked whether this assumption 
is appropriate.

All our results are rather insensitive to the exact value 
of distance $R_{out}$ at which atmosphere finally merges with the 
nebula as long as $R_{out}\gtrsim R_B$. This is because 
atmospheric pressure outside $R_B$ is not very 
different from $P_0$ --- planetary gravity cannot strongly 
perturb the pressure, while temperature is not very different 
from $T_0$: in the case of envelopes having outer 
convective zone deviation of $T$ from $T_0$ is directly  
related to the deviation of $P$ from $P_0$ by the condition of 
adiabaticity  and, consequently, must be small outside of Bondi sphere. 
In the case of envelopes having outer radiative zone temperature
gradient is subadiabatic meaning that temperature deviation 
outside $R_B$ is even smaller than in the convective case. 
As a result, atmospheric pressure and temperature are not very 
different from their nebular values already at $R_B$
and the exact location where $P$ and $T$ 
closely  match $P_0$ and $T_0$ is not very important. 

Our use of opacity in the form (\ref{eq:opacity}) may seem 
an oversimplification compared to other treatments (e.g. Mizuno 
1980; Ikoma \etal 2000) which employ realistic opacity tables. 
However, as mentioned previously, we expect $\kappa$ in the 
outer part of the envelope to be dominated by dust and this 
typically (for $\beta\approx 1-2$) leads to convection inward 
from $R_a$ obviating the need to know the gaseous opacity 
behavior at high temperatures and pressures. Some effect on the 
envelope mass and critical core mass might come from the change 
of equation of state caused by dissociation and ionization in 
the deep interior of the atmosphere, which e.g. might lead to 
the appearance of radiative regions at large depth.
However, since atmospheric mass budget is dominated by the outer 
part of the envelope at $r\sim R_a$, change of $\kappa$  
or $\gamma$ deeper down hopefully would not strongly affect 
the value of $M_{env}$.

Thus, we expect our simple analytical treatment to
provide robust qualitative picture of the envelope 
structure and its dependence upon local conditions in the 
protoplanetary disk, and yield reasonable quantitative estimates 
of $M_{env}$ and $M_{cr}$.


\section{Conclusions.}
\label{sect:concl}

We investigated steady-state structure of the atmosphere
around protoplanetary core immersed in the gaseous disk. Our 
major results can be summarized as follows.

Atmospheres split into those having outer convective  
or outer radiative (almost isothermal) zone. The former have entropy of the 
interior equal to the entropy of the surrounding nebular gas; 
the latter have interior entropy which is much lower 
than the nebular entropy, owing to the decoupling 
provided by the outer radiative region. 
Type of atmosphere around a given core is determined by the
relationships between the Bondi radius $R_B$, photon mean free path
$\lambda$, and luminosity radius $R_L$. If envelope is optically 
thick at the Bondi radius ($R_B\gg \lambda$) atmospheric type
is set only by the value of the dimensionless parameter $R_B\lambda/R_L^2$,
inverse of which has a meaning of radiative temperature gradient 
far from the core: outer envelope is convective when 
$R_B\lambda/R_L^2\lesssim 1$, while when this parameter 
is $\gtrsim 1$ atmosphere has an outer radiative zone.
For the conditions typical in the protoplanetary disks 
(such as MMSN) cores having atmospheres of the first kind 
are common in the terrestrial planet region; in the region of 
giant planets atmospheres have outer radiative zone.
Structure of the atmospheric interior is determined by the 
dependence of opacity on gas temperature and pressure; 
it becomes especially simple (convective as soon as temperature 
starts to appreciably vary with depth) if opacity 
is dominated by small dust grains. 

In the terrestrial region critical core mass for 
nucleated instability $M_{cr}$ depends only 
on the local values of nebular gas density and temperature. 
In the region of giant planets $M_{cr}$ is insensitive to 
either $\rho_0$ or $T_0$ whenever opacity in the outer radiative zone
is independent of the gas pressure, because outer radiative zone  
decouples inner parts of the envelope from the nebular gas 
(irrespective of whether the atmospheric interior is 
radiative or convective); at the same time $M_{cr}$ is a strong 
function of accretion luminosity, opacity, and mean molecular 
weight. Critical core mass 
varies as a function of distance from the Sun because
of the variation of $\rho_0$ and $T_0$ in the terrestrial 
planet region and because of the variation of planetesimal 
accretion luminosity and photon mean free path 
in the region of giant planets. 
Typical value of $M_{cr}$ is several tens of $M_\oplus$ 
in the giant planet region ($30-50$ M$_\oplus$ at $5$ AU) if planetesimal 
accretion was fast enough to account for the core formation 
prior to the gas dissipation. 
Close to the Sun, at $a\sim 0.1$ AU, $M_{cr}\approx 5$ 
M$_\oplus$ independent of the planetesimal accretion rate. This 
makes in situ formation of ``hot Jupiters'' by nucleated 
instability onto the locally formed protoplanetary cores very 
unlikely.

Results of this study are also important for understanding the 
ancient atmospheres of terrestrial planets, which must have been 
significantly depleted.

\acknowledgements 

I am grateful for hospitality to Kavli Institute for 
Theoretical Physics where
part of this work has been done. Useful discussions 
with Doug Lin, Peter Goldreich, and Bruce Draine 
are thankfully acknowledged. 
Author is a Frank and Peggy Taplin Member at the IAS;
he is also supported by the 
W. M. Keck Foundation and NSF grants PHY-007092, PHY99-0794.

\appendix


\section{A. Summary of the core accretion rates.}
\label{ap:acc_rate}

Following Rafikov (2003b) we take the core accretion rate to be
\begin{eqnarray}
\dot M=\Omega \Sigma_p R_c R_H\theta,
\label{eq:m_dot}
\end{eqnarray}
where $\theta$ is a parameter set by a particular mode of 
accretion. In this work we consider the following three 
important accretion regimes. 

First one is characterized by $\theta\approx p\ll 1$ (see definition 
[\ref{eq:p_def}]) and assumes that core accretes planetesimals 
at a rate set by the core's geometric cross-section 
$\sim R_c^2$. This regime is valid when the random epicyclic 
velocities of planetesimals are larger than the escape speed
from the core's surface and gravitational focusing is weak. 
We call this regime {\it slow accretion}.
It may occur in planetesimal disks after cores have reached 
isolation mass by oligarchic growth (Chambers 2001; Goldreich 
\etal 2004). 

Second regime of {\it intermediate accretion} takes place when 
random velocities of planetesimals are of the order of 
shear velocity across the Hill radius of the core $\Omega R_H$. 
In this case
gravitational focusing strongly increases accretion cross-section
above its geometric value, and $\theta\approx 1$.
Note that this case corresponds to the boundary between the shear- 
and dispersion-dominated dynamical regimes (Stewart \& Ida 2000); it
also assumes vertical scaleheight of the planetesimal disk to 
be $\sim R_H$. This regime may occur during the oligarchic growth
of protoplanetary embryos  by accretion of large planetesimals 
(Kokubo \& Ida 1998; Rafikov 2003b).

Finally, third regime is realized when random 
velocities of planetesimals are so low that planetesimal
disk is geometrically very thin and essentially two-dimensional. 
For that one needs random 
velocities to be smaller than $p^{1/2}\Omega R_H$, which leads to
a {\it rapid accretion} with $\theta\approx p^{-1/2}$. Such 
dynamically ``cold'' planetesimal populations can occur even in
the presence of massive cores provided that some dissipative 
process such as gas drag can effectively damp planetesimal 
velocities. This can happen if fragmentation of large
planetesimals in collisions at high velocities (induced by the
gravity of protoplanetary cores) is capable of channeling a 
significant amount of mass initially locked up in large 
planetesimals into small bodies; see Rafikov (2003a,b) 
for details of this scenario. 

It is likely that each of these accretion regimes can
take place during some stage of protoplanetary growth,
e.g. starting with intermediate, switching to rapid
(or some mixture of rapid and intermediate, depending on the 
fragmentation timescale, see Rafikov [2003b]), and, possibly, 
ending with a slow accretion phase. Typical accretion 
timescale $\tau_{acc}$ in each regime is
\begin{eqnarray}
\tau_{acc}\equiv\frac{M_c}{\dot M}=
\frac{M_c}{\Omega \Sigma_p R_c R_H}\theta^{-1}
\approx \left(\frac{M_c}{M_\oplus}\right)^{1/3}
\left\{
\begin{array}{l}
3\times 10^{10}~\mbox{yr}~a_{10}^{3},~~~~\mbox{slow},\\
1.4\times 10^7~\mbox{yr}~a_{10}^{2},~~~\mbox{intermediate}, \\
3\times 10^5~\mbox{yr}~a_{10}^{3/2},~~~~\mbox{fast}.
\end{array}
\right.
\label{eq:t_acc}
\end{eqnarray}
Note that in all three  
regimes listed here protoplanetary growth proceeds as 
$M_c\propto \tau^3$, where $\tau$ is the time.


\section{B. Calculation of the envelope mass.}
\label{ap:mass}

As we mentioned in \S \ref{subsect:crit_mass}, because  gas 
density drops exponentially outside $R_a$, most of the atmospheric
mass is confined within this radius and we may replace $R_B$ with $R_a$
in definition (\ref{eq:m_env}).
Using (\ref{eq:rad_zero_t}) and (\ref{eq:rad_zero_p}) for envelopes 
with radiative interior, and (\ref{eq:T_conv}),
(\ref{eq:P_conv}), and (\ref{eq:conv_zero}) for envelopes with 
convective interior we arrive at equation (\ref{eq:mass_env}),
in which $\Psi_1$ is defined by 
\begin{eqnarray}
\Psi_1(u,w,\zeta)\equiv C w^\zeta\left(\frac{R_a}{R_B}\right)^{3-\zeta}
\int\limits_0^1 z^{2}
\left(\frac{1}{z}-1\right)^\zeta dz,
\label{eq:theta_def}
\end{eqnarray}
and for envelopes having radiative interior 
($\nabla_0<\nabla_{ad}$)
\begin{eqnarray}
C=\left(\frac{4\nabla_0}{3}
\right)^{1/(1+\alpha)},~~~u=\xi,~~~w=\nabla_0,~~~\zeta=\frac{1}{\nabla_0}-1,
\label{eq:rad_case}
\end{eqnarray}
while in the case of envelopes having convective interior 
($\nabla_0>\nabla_{ad}$)
\begin{eqnarray}
C=\left(1-\frac{\nabla_{ad}}{\nabla_0}
\right)^{1/(4-\beta)}\left(\frac{4}{3}\frac{\nabla_0
\nabla_{ad}}{\nabla_0-\nabla_{ad}}
\right)^{1/(1+\alpha)},~~~u=1,~~~ w=\nabla_{ad}
\left(1-\frac{\nabla_{ad}}{\nabla_0}\right)^{1/(4-\beta)},
~~~\zeta=\frac{1}{\gamma-1}.
\label{eq:conv_case}
\end{eqnarray}
Constant $\zeta$ is 
a power law index of density dependence on $1/r$,
see equations (\ref{eq:rad_zero_t}), (\ref{eq:rad_zero_p}) 
and (\ref{eq:conv_sol}). 

Whenever $\zeta<3$ integral in (\ref{eq:theta_def}) is 
dominated by the contribution from $r\sim R_a$, 
in which case $M_{env}$ depends 
only weakly on the lower integration limit $R_{c}\ll R_a$
(this is why we set $R_c/R_B$ to 0 in  
[\ref{eq:theta_def}]). Radiative envelopes 
with constant opacity ($\alpha=\beta=0$) having
$\nabla_0=1/4$ and $\zeta=3$ contain equal 
amount of mass per every decade in radius; in this case formula 
(\ref{eq:mass_env}) still describes the behavior of 
$M_{env}$ but with $\Psi_1\sim \ln(R_a/R_c)$
replacing (\ref{eq:theta_def}). This
can considerably increase the envelope mass since $R_a\gg R_c$ . 
Condition $\zeta<3$ is always
satisfied for dynamically stable convective
envelopes which ought to have $\gamma>4/3$. In the 
radiative case one needs $\nabla_0>1/4$ for $\zeta<3$; 
whenever this is not fulfilled the envelope mass is dominated 
by the innermost part of the atmosphere near $R_c$, and 
$M_{env}$ does depend on $R_c$.  We do not consider 
radiative atmospheres having $\zeta>3$ deep in the 
envelope in this study.


\section{C. Thermal timescale of the envelope.}
\label{ap:thermal_time}

We calculate the thermal (or Kelvin-Helmholtz) time for the 
atmosphere with the outer radiative zone 
(see \S \ref{sect:solutions}) and convective interior 
($\alpha=0,~\beta=1$, $\nabla_0=1/3>\nabla_{ad}$) with 
$\gamma=7/5$. We define thermal time as 
$\tau_{th}\equiv |E_{tot}|/L$, where $E_{tot}=E_{th}+E_{gr}$
is the total energy contained within $R_a$ --- sum of the 
thermal and gravitational energy of the 
envelope. It is easy to verify 
that $E_{th}\sim |E_{gr}|$, meaning
that $E_{tot}\sim E_{gr}$ as well.
We calculate $E_{gr}$ using 
(\ref{eq:T_conv}), (\ref{eq:P_conv}), and 
(\ref{eq:conv_zero}):
\begin{eqnarray}
&& E_{gr}=-\int\limits_{R_c}^{R_a}G\frac{M_c}{r}\times 4\pi 
\rho(r)r^2dr\nonumber\\
&& \approx -4\pi P_0 R_B^3\left(\frac{R_B}{R_c}
\right)^{\frac{3-2\gamma}{\gamma-1}}\left(
\frac{4}{3}\frac{\nabla_{ad}\nabla_0}{\nabla_0-\nabla_{ad}}
\frac{R_B \lambda}{R_L^2}\right)^{\frac{1}{1+\alpha}}
\frac{\gamma-1}{3-2\gamma}\left(1-\frac{\nabla_{ad}}
{\nabla_0}\right)^{\frac{1}{\nabla_{ad}(4-\beta)}}
\nabla_{ad}^{\frac{1}{\gamma-1}}.
\label{eq:grav_energy}
\end{eqnarray}
This expression is valid for envelopes with convective 
interiors whenever $\gamma<3/2$, in which case 
gravitational energy budget is dominated by the {\it innermost} part of 
the envelope, near the core surface (see Harris 1978); this is why 
$E_{gr}$ in 
(\ref{eq:grav_energy}) explicitly depends on $R_c$. 
This would be different for atmospheres with convective 
interiors having $\gamma>3/2$ --- then the energy content
is dominated by the {\it outer} parts of the envelope\footnote{The 
latter is also true for envelopes with radiative interiors having
$\nabla_0>1/3$ (which can exist only for $\gamma>3/2$); most of the 
energy in the radiative envelopes with $\nabla_0<1/3$ is near the 
core.} 
and is much smaller than that given by (\ref{eq:grav_energy}).
Energy is also small for envelopes having outer convective 
zone  --- their energy is low because of the high interior 
entropy and associated low density. In a sense, the specific 
estimate (\ref{eq:grav_energy}) of $E_{gr}$ sets an upper limit 
on $\tau_{th}$ and the degree of envelope's deviations from the 
steady state. 

Using our
adopted values of $\gamma, \nabla_0, \nabla_{ad}$ and
luminosity (\ref{eq:luminosity}), (\ref{eq:m_dot}) we 
find that
\begin{eqnarray}
\tau_{th}\approx\left(\frac{M_c}{M_\oplus}\right)^{5/3}
\left(\frac{0.1~\mbox{cm$^2$g$^{-1}$}}{\kappa_0}\right)
\left\{
\begin{array}{l}
10^{9}~\mbox{yr}~a_{10}^{23/4},~~~~\mbox{slow},\\
10^3~\mbox{yr}~a_{10}^{15/4},~~~~\mbox{intermediate}, \\
1~\mbox{yr}~a_{10}^{11/4},~~~~~~~\mbox{fast}.
\end{array}
\right.
\label{eq:tau_th}
\end{eqnarray}
Comparing this with (\ref{eq:t_acc}) one can see that typically 
$\tau_{th}\ll \tau_{acc}$ justifying our quasi-static 
approximation to the treatment of the envelope structure. 
Massive embryos ($M_c\gtrsim 10~M_\oplus$) 
accreting in the slow regime at $a>10$ AU should have this 
condition violated and this may pertain to the development
of nucleated instability in the region of giant planets, see
\S \ref{subsect:core_mass}.

At a given location in the nebula the energy stored 
in the envelope depends only on $M_c$. Since $M_c$ changes 
due to planetesimal accretion, envelope energy
should also vary in time giving rise to additional 
luminosity $L_g$ caused by {\it gas accretion}: 
\begin{eqnarray}
L_g\equiv\frac{\partial |E_{tot}|}{\partial \tau}=
\frac{\partial |E_{tot}|}{\partial M_c}\dot M\sim 
\frac{|E_{tot}|}{\tau_{acc}}.
\end{eqnarray}
Using the definition of $\tau_{th}$ we then find 
that $L_g/L\sim \tau_{th}/\tau_{acc}$. Thus, whenever the 
quasi-stationary approximation (i.e. $\tau_{th}\ll 
\tau_{acc}$) is valid, gas accretion luminosity $L_g$ is 
{\it small} compared to the core accretion luminosity 
$L$, and we can safely neglect it. Mass accretion rate of gas
$\dot M_{env}\sim \dot M(M_{env}/M_c)$ is always smaller
than planetesimal mass accretion rate $\dot M$ if core
is subcritical, i.e. $M_{env}\lesssim M_c$.


\section{D. Efficiency of convective transport.}
\label{ap:conv_efficiency}

Our use of equation (\ref{eq:polytrope}) relies on the 
assumption of convection so efficient that even 
infinitesimal deviation of temperature gradient from 
$\nabla_{ad}$ is enough to transport the 
energy flux produced at the core surface. If this is not 
the case one has to use mixing-length theory 
(Kippenhahn \& Weigert 1990) to determine the value of 
$\nabla$; here we check if this is ever necessary. 
Following Kippenhahn \& Weigert (1990) we introduce 
\begin{eqnarray}
&&
x\equiv\nabla-\nabla_{ad},~~~W\equiv\nabla_{rad}-\nabla_{ad},~~~
U\equiv \frac{6\sqrt{2}\nabla_{ad}}{ \eta^2}\frac{\sigma T^4}
{P\kappa\rho}
\left(\frac{r^2}{GM_cH_p^3}\right)^{1/2},\nonumber\\
&&
\nabla_{rad}\equiv\frac{3}{16\pi\sigma G}
\frac{\kappa L P}{M_c T^4},~~~H_p\equiv
\frac{\partial r}{\partial\ln P},
\label{eq:defs}
\end{eqnarray}
where $\eta\sim 1$ is a mixing length parameter, $H_p$ is the pressure 
scaleheight, and $x$ is a deviation of temperature gradient 
from $\nabla_{ad}$, which has to be much smaller than unity
for (\ref{eq:polytrope}) to apply. Value of $x$ has to
be obtained from the following equation (Kippenhahn \& Weigert 
1990):
\begin{eqnarray}
\left(\sqrt{x+U^2}-U\right)^3=\frac{8}{9}U(W-x).
\label{eq:x_eq}
\end{eqnarray}
It is clear from this equation that $x\ll 1$ whenever 
$U\ll 1$, and $UW\ll 1$, since under these 
assumptions $x\sim (UW)^{2/3}$. Thus, we look for 
conditions under which $U\ll 1, UW\ll 1$. 

In light of the results of \S \ref{subsect:convection} \& 
\ref{sect:solutions_inner} we describe the profiles of 
temperature, density, and pressure in the convective part 
of envelope by
\begin{eqnarray}
T=T_b\Phi,~~~\rho=\rho_b\Phi^{1/(\gamma-1)},
~~~P=P_b\Phi^{\gamma/(\gamma-1)},~~~
\Phi\equiv 1+\nabla_{ad}\frac{T_0}{T_b}
\left(\frac{R_B}{r}-\frac{R_B}{r_b}\right),
\label{eq:conv_struct}
\end{eqnarray}
where $T_b, \rho_b, P_b$ are values of temperature, density 
and pressure at the boundary of the convective zone $r_b$. In the
case of an atmosphere with the outer radiative zone studied in 
\S \ref{subsect:convection} one has $r_b=R_a, T_b=T_{conv},
\rho_b=\rho_{conv}, P_b=P_{conv}$, and (\ref{eq:conv_struct})
reduces to (\ref{eq:conv_sol}). In the case of atmosphere with 
the outer convective zone $r_b=R_{out}, T_b=T_0, \rho_b=\rho_0, 
P_b=P_0$ and (\ref{eq:conv_struct}) yields (\ref{eq:conv_sol_out}). 

One expects deviations of $x$ from zero to be most 
pronounced in the outer, low density part of the atmosphere  
which might not be capable of efficient mixing.
This corresponds to $r_b\approx R_a$ in the case of outer radiative 
zone, and $r_b\approx R_B$ in the case of outer convective zone, but
in both cases $\Phi(r_b)\approx 1$. In the atmospheres of first type 
($R_L^2\ll R_B\lambda$) we find for $\alpha=0, \beta=1, 
\gamma=7/5$ using (\ref{eq:T_conv}), (\ref{eq:P_conv}), 
(\ref{eq:R_a}) that
\begin{eqnarray}
&& U(R_a)\approx \frac{6}{\eta^2}\frac{\sigma T_0^4}
{c_0\kappa_0\rho_0^2 GM_c}
\left(\frac{R_L^2}{R_B\lambda}\ln
\frac{R_B\lambda}{R_L^2}\right)^2\nonumber\\
&& \approx 
\eta^{-2}\left(\frac{M_\oplus}{M_c}\right)^{1/3}
\left(\frac{\kappa_0}{0.1~\mbox{cm$^2$g$^{-1}$}}\right)
\left\{
\begin{array}{l}
3\times 10^{-10}~a_{10}^{-19/4},~~~~\mbox{slow},\\
3\times 10^{-4}~a_{10}^{-11/4},~~~~~\mbox{intermediate}, \\
0.2~a_{10}^{-7/4},~~~~~~~~~~~~~~\mbox{fast}.
\end{array}
\right.
\label{eq:U_rad}
\end{eqnarray}
Also $\nabla_{rad}\sim\nabla_{ad}$ at $r\approx R_a$, meaning that 
$W(R_a)\sim 1$ and $UW\sim U(R_a)$. Thus, in the region of 
giant planets ($a\gtrsim 5$ AU)
convection is efficient in the envelopes of protoplanetary
cores accreting planetesimals in the slow or intermediate regime. 
In the case of fast accretion deviations of $\nabla$ from 
$\nabla_{ad}$ at the level of $0.1-1$ might be expected in 
the outermost parts of convective zone, at $r\approx R_a$. 

In the case of envelopes having outer convective zone 
($R_L^2\gg R_B\lambda$) one finds 
for $U(R_B)$ expression similar to (\ref{eq:U_rad}) but without
$R_L^2/R_B\lambda$ terms. Also, 
$\nabla_{rad}\sim R_L^2/R_B\lambda\gg 1$ at $r=R_B$. 
As a result, 
\begin{eqnarray}
&& U(R_B)\approx \frac{0.03}{\eta^2}
\left(\frac{M_c}{M_\oplus}\right)
\left(\frac{0.1~\mbox{cm$^2$g$^{-1}$}}{\kappa_0}\right)
a_1^{15/4},\nonumber\\
&& W(R_B)U(R_B)\approx 0.5
\frac{L}{\rho_0 c_0^3 R_B^2}\approx 
\eta^{-2}
\left(\frac{M_\oplus}{M_c}\right)^{2/3}
\left\{
\begin{array}{l}
10^{-4}~a_{1}^{-1/2},~~~~\mbox{slow},\\
0.03~a_{1}^{1/2},~~~~~~\mbox{intermediate}, \\
0.4~a_{1},~~~~~~~~~~\mbox{fast}.
\end{array}
\right.
\label{eq:U_conv}
\end{eqnarray}
These estimates demonstrate that in  protoplanetary atmospheres 
around the cores at $a\sim 0.1$ AU, which are likely to have
outer convective zones, deviations from purely isentropic
convection may only be important for fast and 
intermediate accretion regimes.

These are the occurrences when the temperature gradient can
go superadiabatic at the outer edge of the convection zone
and one would get a more accurate estimate of $\nabla$ in the
convective regions by actually solving (\ref{eq:x_eq}),
e.g. see Papaloizou \& Terquem (1999).
We do not expect this possible superadiabaticity 
in some localized parts of atmosphere to strongly affect 
our estimates of $M_{env}$ and $M_{cr}$ since deeper down 
in the envelope density increases, convection becomes more 
efficient at transporting the energy, and entropy gradient 
goes to zero.


\end{document}